\documentclass[conference]{IEEEtran}

\usepackage{graphicx}
\usepackage{amssymb,amsmath,latexsym,amsfonts}
\usepackage{epsfig}
\usepackage{comment}
\usepackage{epstopdf}
\usepackage{color}
\usepackage[top=0.9in, bottom=0.9in, left=2.2cm, right=2.2cm]{geometry}
\usepackage{setspace}
\newcommand{\be}{\begin{equation}}
\newcommand{\bel}[1]{\begin{equation}\label{#1}}
\newcommand{\ee}{\end{equation}}
\newcommand{\ba}{\begin{align}}
\newcommand{\ea}{\end{align}}
\newcommand {\N} {{\rm I\kern-2.5pt N}}
\newtheorem{lemma}{\hspace{-0.15in}{\bf Lemma}}
\newtheorem{theorem}{\hspace{-0.15in}{\bf Theorem}}

\newtheorem{definition}{\hspace{-0.15in}{\bf Definition}}

\def\done{\hspace*{\fill} \rule{1.8mm}{2.5mm}}

\newcommand{\SA}{ {\cal A} }
\newcommand{\SB}{ {\cal B} }
\newcommand{\SD}{ {\cal D} }
\newcommand{\SH}{ {\cal H} }
\newcommand {\bx}{\mbox{\boldmath $x$}}
\newcommand {\by}{\mbox{\boldmath $y$}}

\ifCLASSINFOpdf

\else

\fi

\begin{document}
%
\title{Multi-dimensional Network Security Game: \\
{\huge How do attacker and defender battle on parallel targets?}}
\author{\authorblockN{Yuedong Xu$^*$, John C.S. Lui$^{\dag}$}
\authorblockA{$^*$Department of Electronic Engineering, Fudan University, China\\
$^{\dag}$ Department of Computer Science \& Engineering, The Chinese University of Hong Kong, China\\
Email: \{ydxu@fudan.edu.cn, cslui@cse.cuhk.edu.hk\}
}
}

\maketitle

\begin{abstract}

In this paper, we consider a new network security game wherein an
attacker and a defender are battling over ``{\em multiple}'' targets.
This type of game is appropriate to model
many current network security conflicts such as Internet phishing,
mobile malware or network intrusions. In such attacks,
the attacker and the defender need to decide how to allocate resources on each target
so as to maximize his utility within his resource limit.
We model such a multi-dimensional network security game as a 
constrained non-zero sum game. Two security breaching models, the product-form
and the proportion-form, are considered. For each breaching model, 
we prove the existence of a unique Nash equilibrium (NE) based on Rosen's theorem
and propose efficient algorithms to find the NE when the games are strictly concave. 
Furthermore, we show the existence of multiple NEs in the 
product-form breaching model when the strict concavity
does not hold. 
Our study sheds light on the strategic behaviors of the attacker and the 
defender, in particular, 
on how they allocate resources to the targets which have different weights, and 
how their utilities as well as strategies are influenced by the resource constraints. 

\end{abstract}


\IEEEpeerreviewmaketitle

\section{{\bf Introduction}}

\IEEEPARstart{T}{he} economics of network 
security has become a thriving concern in 
fixed line and mobile Internet.  Due to the popularity of e-commerce 
and other online services, 
malicious attacks have evolved into profit driven online crimes in the forms of 
Internet phishing, network intrusion, mobile malware etc. 
Although security defence is essential,
the networking community is still witnessing an increased number 
of global attacks. 
Part of reasons are the economic benefits on performing attacks by hackers
as well as the inadequate protection against the persistent attacks. 
Therefore, economic studies
beyond the technological solutions are vitally important to reveal the 
behaviors of the defenders and the malicious attackers, and 
game theory serves as a well suited mathematical tool
to bring about this fundamental understanding. 
A prominent application of game theory in security is 
intrusion detection where an attacker exploits
system vulnerabilities and a defender monitors the events 
occurring in a network strategically \cite{tifs09}\cite{alpcan}. 
Recent advances of network security games have two features. 
One is called \emph{uncertainty} that incorporates 
incomplete information of players \cite{alpcan}
and stochastic properties of players or environments \cite{zhu2}. 
The other is called \emph{interdependency} in which the actions of players
may affect other players. This type of interactions are sometimes regarded as 
network effects with positive or negative
externality \cite{chuang1,john3,lelarge2}.

In this work, we explore a new type of network security game 
which is characterized by
\emph{multi-dimensional} attacks. We are motivated by three facts. 
Firstly, the effectiveness of attack or defence depends on the amount of resources 
that are used. 
The resource is an abstract representation of manpower, machines, 
technologies, etc. For instance, many resources are needed to 
create malicious websites in phishing attacks, or to camouflage malicious
apps in mobiles, or to recruit zombie machines in DDoS attacks, or to probe server 
vulnerabilities in intrusion attacks. 
However, one needs to note that resource is \emph{not free} for the attacker and the defender.
Secondly, the attacker and the defender usually possess limited resources. 
For instance, the number of active bots that a botmaster can manipulate
is usually constrained to a few thousands \cite{tpds14}. 
Thirdly, the attacker can assaults \emph{multiple} targets for better economic returns. 
These targets may represent different banks in the Internet phishing attack \cite{report2}, 
or different Android apps in mobile malware, or different servers in network 
intrusion attacks. These targets vary in values or importances. 
Attacking (resp. protecting) more targets requires a larger amount 
of resources, which may exceed the resource budget of the attacker (resp. defender). 
As a consequence, the conflicts on multiple targets are conjoined whenever 
the attacker or the defender has limited amount of resources. This transforms the  
decision making in network security issues into 
myopic constrained optimization problems.

We propose a non-zero sum game-theoretic framework to characterize 
the constrained resource allocation between an attacker and
a defender. The utility of the attacker is modeled as the profit, 
which is equivalent to the loss of victims minus the costs of attack resources. 
The utility of the defender is modeled 
as the loss of victims plus the costs of defence resources. Both players aim 
to optimize their individual utilities. 
We express the loss of victims on a target as a product of its weight and 
the security breaching probability. Two breaching models are considered;
one is the product-form of attack and defence efficiencies, the other
is the proportion-form of attack and defence efficiencies.
In our work, we focus on the following questions:
\emph{1) How does a player select targets to attack/defend and how does he 
 allocate resources to heterogenous targets at the NE? 
 2) How do the resource limits of the players influence the NE and their performance 
 at the NE?
}

This work provides important insights into the multi-dimensional network security issues. 
In the \textit{product-form} breaching model, 
both players allocate positive resources to 
the subsets of more valuable targets at the NE. For any two targets protected by the defender, 
he always allocates more resources to one with a higher value. 
While the attacker may allocate more resources to the more important targets, 
or evade the well-protected valuable targets, 
depending on the defender's \emph{relative ineffectiveness of defence (RID)}.  
We also show the existence of multiple NEs that yield different utilities to the players 
in some special scenarios. The attacker and the defender may place positive 
amount of resources to more targets when they acquire larger resource budgets. 
When both players do not possess sufficient amount of resources, anyone of them 
may improve his utility if his resource limit increases. 
However, when the defender always has sufficient amount of resources, the increase 
of attacker's resource limit can lead to an \emph{arm race} in which both players 
may obtain smaller utilities at the NE. 
In the \textit{proportion-form} breaching model, the attacker and the defender allocate 
resources on all the targets. Each player allocates more resources to more valuable 
targets. The resource insufficient player can improve his utility by acquiring more resources. 
Our major contributions are summarized as below: 



- We propose a novel network security game framework 
       that captures the competition on multiple targets simultaneously.

- We present efficient algorithms to find the unique NE when
the objective functions are strictly concave.

- We also show the existence of multiple NEs when 
the objective functions are not strictly concave.

- We provide important insights on how the attacker(s) and the defender allocate
resources to heterogeneous targets under resource constraints, and how the NE(s) 
is (are) influenced by the resource constraints.


The remainder of this paper is as follows. 
Section \ref{sec:model} describes the game model. 
Section \ref{sec:nash} carries out the analysis of the NE as well as the impact of 
resource limits on the NE. 
Section \ref{sec:intrusion} presents a linear intrusion detection game analysis. 
We analyse the NE of the proportion-form breaching model in Section \ref{sec:coupled}.
Section \ref{sec:related} 
surveys the related works and
Section \ref{sec:conclusion} concludes.

\section{{\bf Game Model and Basic Properties}}
\label{sec:model}

In this section, we present a game-theoretic model for network security issues. 
The game contains two players, one being an attacker and the other being a 
defender, they simultaneously compete on multiple targets. We first provide some 
salient features of current network security attacks, and then present our model formulation.

\subsection{Motivation}
\vspace{-0.4cm}

We are motivated by new features of network attacks and defences that are not 
well captured by existing works (e.g. \cite{alpcan} and reference therein). 
Most of state-of-the-art researches focus on the one dimensional strategies 
(i.e., monitoring probability of intrusion, channel access probability or insurance adoption of a node). 
Such game models are insufficient to characterize the modern days security attacks such as 
phishing and mobile malware, etc. Here, we present some salient features of 
network attacks that lead to our game formulation.

\textit{First, the attackers and the defenders are resource constrained.} Resources are defined in a variety of forms. For instance, in the fast-flux phishing attack, the hijacked IP address is one type of resources of the attackers. In a mobile malware attack, the attacker's resources can be the technology and the manpowers used to spoof the security check mechanism of the third-party apps markets. 
In DDos attacks, a botmaster is usually able to control only a few thousands active bots \cite{tpds14}.
Similarly, the defender needs to allocate resources such as technologies and manpowers to detect and remove these attacks. In general, both the attacker and the defender only possess limited resources.

\textit{Second, the efficiencies of attacks and defences depend on how many resources are allocated.} While existing works 
(e.g. references in \cite{alpcan}) assume that the payoffs of the attacker and the defender are determined by whether the target is attacked or defended. 
We take phishing attack as an example. By creating more malicious websites, the phishing attacker is able to seduce 
more users and to perform more persistent attacks. If the defender allocates more resources to perform proactive detection, 
more malicious sites will be ferreted out in zero-day, and the attack time window will be reduced. 
Similarly, if more efforts are spent to create 
malicious Android apps, the attacker can carry out more effective camouflage, thus gaining more profits through stealing 
private information or sending premium SMS imperceptibly. As a countermeasure, the defender will install these apps on his cloud and 
examine their suspicious events for a certain amount of time.

\textit{Last but not least, the attacker and the defender battle not on a single, but rather, multiple targets.} 
Attackers are profit-driven. They are inclined to attack many targets in parallel. The targets are specified as 
different E-banks in phishing, different apps in mobile malware attacks and
different servers in network intrusions. Note that the targets vary in their valuations, 
so the attacker and the defender may allocate different amount of resources to them. 
To attack (resp. protect) multiple targets, more resources are required. 
How to perform parallel attacks becomes a challenging problem 
when players have certain resource limits. All these motivate our study on 
the strategic allocation of limited resources by the players on multiple targets simultaneously.

\subsection{Models}

Let us start with the basic security game which consists of two players, 
an attacker $\SA$ and a defender $\SD$. The attacker
launches attacks on $N$ targets (or ``battlefields'' interchangeably) 
which we denote as $\SB=\{B_1,{\cdots},B_N\}$. The target $B_i$ is associated with a 
weight $w_i$ ($i=1,{\cdots},N$). 
When $w_i {>} w_j$, $B_i$ is more valuable than $B_j$. 
Without loss of generality, we rank all targets from 1 to $N$
in the descending order of their weights (i.e. $w_i{>}w_j$ if $i{<}j$).

Attacking a target may consume some resources such as 
manpower to design malware, social engineering 
techniques to camouflage them, or dedicate many compromised 
machines for attacks. Defending a target needs manpower, investment 
in technology, and computing facilities etc. 
Here, we monetarize different types of resources. 
Let $c$ be the price of per-unit of $\SA$'s resources, and let $\hat{c}$ be
that of $\SD$'s resources. We next define two important terms that form 
the utilities of the attacker and the defender. 

\begin{itemize}

\item \textit{Attack efficiency.} Let $x_i$ be the amount of resources spent by $\SA$ on $B_i$, 
and let $f(x_i)$ be the corresponding attack efficiency on target 
$B_i$. Here, $f(\cdot)$ reflects the ability of the attacker to 
intrude a system, or to camouflage the malware, etc. 
We assume that $f(x_i)$ is a differentiable, strictly increasing and 
concave function with respect to (w.r.t.) $x_i$. The concavity means that 
the increment of attack efficiency decreases when $\SA$ further increases $x_i$. 
Without loss of generality, we let $f(0)=0$ and $0\leq f(x_i)\leq 1$. 

\item \textit{Defence efficiency.} Denote $y_i$ as the resources that $\SD$ uses to detect and remove the 
attacks on target $B_i$. Let $g(y_i)$ be the defence efficiency when $\SD$ allocates $y_i$ to $B_i$.
We assume that $g(y_i)$ is a differentiable, strictly increasing and concave function of $y_i$ with
$g(0)=0$ and  $0\leq g(y_i)\leq 1$. For the sake of convenience, we define 
a complementary function $\tilde{g}(y_i)$, the defence inefficiency, 
which has $\tilde{g}(y_i)=1-g(y_i)$. Then, $g(\cdot)$ is a decreasing and convex function.
\end{itemize}

It is very difficult to capture the loss of victims (also the revenue of the attacker) 
due to the obscure interaction between the attack efficiency of $\SA$ and the defence efficiency of $\SD$. 
Here, we formulate two simplified \textit{breaching} models, one is named a ``product-form'' model 
and the other is named a ``proportion-form'' model. Denote by $p_i$ the breaching probability of 
target $B_i$. Then, there exist
\begin{itemize}

\item \textit{Product-form model:} $p_i = f(x_i)\tilde{g}(y_i)$;

\item \textit{Proportion-form model:} $p_i=\frac{f(x_i)}{f(x_i) + g(y_i)}$. 

\end{itemize}
In the product-form model, the change of attack (resp. defence) efficiency causes a linear change of breaching probability.  
For mobile phishing attacks, the defence efficiency can be regarded as the probability of detecting malware, and the 
attack efficiency represents the ratio of victims defrauded by the attacker. Then, the breaching probability can be taken
as a product of attack efficiency and defence inefficiency. A classic example of the product-form model
is the matrix-form intrusion detection game where
$f(x_i)$ and $g(y_i)$ are linear functions \cite{alpcan}. The attack efficiency denotes the probability of performing an attack and 
the defence efficiency denotes the probability of performing a detection action. 
In reality, the resources of the attacker and the defender have a coupled effect on the security of a target. 
The increase of attack efficiency might not yield a linearly augmented breaching probability. 
However, it is very difficult to quantify their coupling. Here, we present a proportion-form breaching model that 
generalizes the cyber-security competition in \cite{ejor} and the DDoS attacks on a single target 
in \cite{sigmetrics12}. The breaching probability increases with the attack efficiency, while at a 
shrinking speed.

In practice, both $\SA$ and $\SD$ have limited
resource budgets which we denote by $X_\SA$ and $Y_\SD$ respectively, 
with $0 < X_\SA, Y_\SD < \infty$. Our focus is to unravel the allocation strategies of the players on multiple targets
with the consideration of resource limits. To achieve this goal, we make the following assumption
on the attack and defence efficiencies. 

\noindent \textit{Assumption:} $\lim_{x_i\rightarrow \infty}f(x_i)=1$ and $\lim_{y_i\rightarrow \infty}{g}(y_i)=1$ in the product-form model if not mentioned explicitly.

Late on, we consider the linear $f(x_i)$ and $g(y_i)$ that generalize intrusion detection game to multiple targets.
As a consequence of attacking $B_i$, 
$\SA$ receives an expected revenue of $w_ip_i$.
Let $U_\SA$ be the aggregate profit of $\SA$ on all the $N$ 
targets. We have $U_\SA = \sum\nolimits_{i=1}^N w_ip_i {-} c\sum\nolimits_{i=1}^N x_i.$
The attacker $\SA$ is usually profit driven and is assumed to be risk-neutral. 
His purpose is to maximize $U_\SA$ under the resource cap $X_\SA$. 
Then, the constrained resource allocation problem is expressed as
{
\setlength{\abovedisplayskip}{0.1cm}
\setlength{\belowdisplayskip}{0.15cm}
\begin{eqnarray}
&\max_{\{x_i\}_{i=1}^N}& U_\SA  \nonumber\\
&\textrm{subject to}& \sum\nolimits_{i=1}^Nx_i \leq X_\SA.
\label{eq:uattacker}
\end{eqnarray}
}
The defender $\SD$'s objective is to minimize the revenue of 
the attacker $\SA$ with the 
consideration of his resource budget.
Let $U_\SD$ be the {\em disutility} of $\SD$ given by 
$U_\SD = -\sum\nolimits_{i=1}^N w_i p_i {-} \hat{c}\sum\nolimits_{i=1}^N y_i.$ 
When $\hat{c}$ (resp. $c$) is 0, $\SD$ (resp. $\SA$) has a use-it-or-lose-it cost structure such that 
he will utilize {\em all} his resources.
The resource allocation problem of $\SD$ can be formulated as:
{
\setlength{\abovedisplayskip}{0.1cm}
\setlength{\belowdisplayskip}{0.15cm}
\begin{eqnarray}
&\max_{\{y_i\}_{i=1}^N}& U_\SD \nonumber\\
&\textrm{subject to}& \sum\nolimits_{i=1}^N y_i \leq Y_\SD.
\label{eq:udefender}
\end{eqnarray}
}
Noticing that $\SA$ and $\SD$ have conflicting objectives, 
we model the resource allocation problem as a two-player non-cooperative game 
and we denote it as $\mathbf{G}$.
Let $\SH$ be a convex hull expressed as $\{(x_i,y_i)|x_i\geq 0, y_i\geq 0, 
\sum\nolimits_{i=1}^{N}x_i\leq X_\SA, \sum\nolimits_{i=1}^{N}y_i\leq Y_\SD\}$.
In what follows, we define a set of concepts for the game. 

\begin{definition}
Nash Equilibrium: 
Let $\bx=(x_1,\cdots,x_N)$ and $\by=(y_1,\cdots,y_N)$ be 
the feasible resource allocations by $\SA$ and $\SD$ 
in the convex hull $\SH$ respectively.
An allocation profile $S=\{\bx^*, \by^*\}$ is a Nash equilibrium (NE) if 
$U_\SA(\bx^*, \by^*) \geq U_\SA(\bx, \by^*)$
and $U_\SD(\bx^*, \by^*) \geq U_\SD(\bx^*, \by)$ 
for any $\bx \neq \bx^*$ and $\by\neq \by^*$.
\label{def:ne}
\end{definition}

\begin{definition} \cite{rosen}
\textit{(Concave game)} A game is called \textit{\textbf{concave}} if each player $i$ 
chooses a real quantity in a convex set to maximize his utility $u_i(x_i,\bx_{-i})$
where $u_i(x_i,\bx_{-i})$ is concave in $x_i$. 
\label{def:concavegame}
\end{definition}

\begin{theorem}\cite{rosen} \textit{(Existence and Uniqueness)}
A concave game has a NE. 
Let $M$ be a $n{\times} n$ matrix function in which $M_{ij}{=}\varphi_i\frac{\partial^2 u_i}{\partial x_i \partial x_j}$,
for some constant choices of $\varphi_i{>}0$. If $M{+}M^T$ is strictly negative definite, then the NE is unique.
\label{theorem:concavegame}
\end{theorem}

\vspace{-0.3cm}

\begin{theorem}
The multi-dimensional security game $\mathbf{G}$ has a unique NE for 
the product-form breaching model if the attack and defence efficiencies are strictly concave, and
for the proportion-form breaching model.
\label{lemma:existence}
\end{theorem}
All the proofs in this work can be found in the appendix.

\section{{\bf Nash Equilibrium and Influence of Resource Limits for Product-form Model}} 
\label{sec:nash}

In this section, we propose an algorithm to find the NE and present its properties.
Furthermore, we analyze how the resource limits $X_\SD$ and $Y_\SD$ influence the allocation 
strategies of the attacker and the defender.

\subsection{Solving NE for the Generalized Game}

In the previous section, we have shown the existence of a
unique NE in the multi-dimensional security game {\bf G}.
However, we have not stated how to derive the NE, which is nontrivial in fact. 
Define $(\bx^*,\by^*)$ as the NE of {\bf G}.
We show that $(\bx^*,\by^*)$ has the following property.

\begin{theorem}
There exist non-negative variables $\lambda$ and $\rho$ such that
{
\setlength{\abovedisplayskip}{3pt}
\setlength{\belowdisplayskip}{3pt}
\begin{eqnarray}
-w_if(x_i^*)\tilde{g}'(y_i^*) - \hat{c}  \;\; \left\{\begin{matrix}
\;=\rho \;\; &&\textrm{ if } \;\; y_i^* > 0 \\
\;\leq \rho \;\; &&\textrm{ if } \;\; y_i^* = 0 
\end{matrix}\right., 
\label{eq:theorem2_no1}
\end{eqnarray}
\begin{eqnarray}
w_i f'(x_i^*)\tilde{g}(y_i^*) - c \;\; \left\{\begin{matrix}
\;=\lambda \;\; &&\textrm{ if } \;\; x_i^* > 0 \\
\;\leq \lambda \;\; &&\textrm{ if } \;\; x_i^* = 0 
\end{matrix}\right.,
\label{eq:theorem2_no2}
\end{eqnarray}
}
\noindent where
{
\setlength{\abovedisplayskip}{3pt}
\setlength{\belowdisplayskip}{3pt}
\begin{eqnarray}
&&\!\!\!\!\!\!\!\!\!\!\left\{\begin{matrix}
\lambda\geq 0 \;\; &&\textrm{ if } \;\; \sum\nolimits_{i=1}^N x_i^* = X_\SA\\
\lambda = 0 \;\; &&\textrm{ if } \;\; \sum\nolimits_{i=1}^N x_i^* < X_\SA
\end{matrix}\right. \;\;\; \textrm{ and }  
\label{eq:slack1}\\
&&\!\!\!\!\!\!\!\!\!\!\left\{\begin{matrix}
\rho\geq 0 \;\; &&\textrm{ if } \;\; \sum\nolimits_{i=1}^N y_i^* = Y_\SD\\
\rho = 0 \;\; &&\textrm{ if } \;\; \sum\nolimits_{i=1}^N y_i^* < Y_\SD
\end{matrix}\right..
\label{eq:slack2}
\end{eqnarray}
}
\label{theorem:ne_condition1}
\end{theorem}

Herein, $\lambda$ and $\rho$ 
are viewed as shadow prices of violating the resource limits.  From 
Theorem \ref{theorem:ne_condition1}, one can 
see that $x_i^*$ and $y_i^*$ may take on 0, 
which occurs when $\SA$ or $\SD$ decides not to attack or defend target $B_i$. 
Our main question here is that given $X_\SA$ and $Y_\SD$, 
how $\lambda$ and $\rho$ are solved at the NE? 
Before answering this question, we state
the sets of targets with positive resources of $\SA$ and $\SD$ at the NE.

\begin{lemma} 
Let $K_\SA$ be the number of targets with positive resources of $\SA$,
and $K_\SD$ be that with positive resources of $\SD$ at the NE. We have 
i) the set of targets being attacked is $\{B_1,\cdots, B_{K_\SA}\}$ and 
the set of targets being defended is $\{B_1,\cdots, B_{K_\SD}\}$;
ii) $K_\SA \geq K_\SD$.
\label{lemma:attackdefendsets}
\end{lemma}

\noindent
{\bf Remark:}
The utility of
Lemma \ref{lemma:attackdefendsets} is that it {\em greatly reduces} the 
space of searching $K_\SD$ and $K_\SA$, which
is essential for us to compute the values 
of $\lambda$, $\rho$, $x_i^*$ and $y_i^*$ at the NE. 
In fact, we only need to 
test at most $(N{+}1)(N{+}2)/2$ possible sets of targets. 
Define two inverse functions 
$h_\SD(\cdot):=\{\tilde{g}'\}^{-1}(\cdot)$ and $h_\SA(\cdot):=\{f'\}^{-1}(\cdot)$. At the NE, 
the resources used by $\SA$ and $\SD$ on a target are given by
\begin{eqnarray}
x^*_i \!\!&=&\!\! \left\{\begin{matrix}
\;h_\SA(\frac{c{+}\lambda}{w_i\tilde{g}(y_i^*(\lambda,\rho))}) \;\; && \forall \;\; i\leq K_\SD  \\
\;h_\SA(\frac{c{+}\lambda}{w_i\tilde{g}(0)}) \;\; &&\forall \;\; K_\SD{<}i{\leq}K_\SA \\
\; 0 \;\; && \forall \;\; i > K_\SA
\end{matrix}\right.,
\label{eq:xy_ne1}\\
y^*_i \!\!&=&\!\! \left\{\begin{matrix}
\;h_\SD(\frac{-(\rho+\hat{c})}{w_if(x_i^*(\lambda,\rho))}) \;\; && \forall \;\; i\leq K_\SD  \\
\;0 \;\; &&\forall \;\; i > K_\SD 
\end{matrix}\right..
\label{eq:xy_ne2}
\end{eqnarray}
In what follows, we 
define a set of notations w.r.t. the total resources 
(denoted as Tot\_Res) 
used by both players at the NE in Table \ref{table:variables1}.
The pair $(X^{suf}_\SA, Y^{suf}_\SD)$ denote the 
\emph{sufficient} amount of resources needed 
by $\SA$ and $\SD$ when $\lambda$ and $\rho$ are both 
0. If both $X_\SA {>} X^{suf}_\SA$ and $Y_\SD {>} Y^{suf}_\SD$ hold, $\SA$ and $\SD$ have 
some unused resources at the NE. Then, the strategies of $\SA$ and
$\SD$ on one target are independent of the other targets. 
We can partition
the plane of $(X_\SA, Y_\SD)$ into four domains: $\mathbf{D_1}$) $X_\SA {\geq} X^{suf}_\SA$ and $Y_\SD {\geq} Y^{suf}_\SD$; $\mathbf{D_2}$) $X_\SA {<} X^{suf}_\SA$ and $Y_\SD {\geq} \hat{Y}^{suf}_\SD$;
$\mathbf{D_3}$) $X_\SA {\geq} \hat{X}^{suf}_\SA$ and $Y_\SD {<} Y^{suf}_\SD$; $\mathbf{D_4}$) none of the above.
If $(X_\SA, Y_\SD)\in D_1$,
the consumed resources of $\SA$ and $\SD$ at the NE are $X_A^{suf}$ and $Y_D^{suf}$ respectively.
If $(X_\SA, Y_\SD)\in D_2$, the resources of $\SA$ are insufficient. 
Then, $\SA$ uses $X_\SA$ resources and $\SD$ uses $\hat{Y}^{suf}_\SD$ at the NE.
If $(X_\SA, Y_\SD)\in D_3$, the resources of $\SD$ are insufficient. Then, 
$\SA$ uses $\hat{X}^{suf}_\SA$ resources and $\SD$ uses $Y_\SD$ at the NE.
If $(X_\SA, Y_\SD)\in D_4$, $\SA$ uses $X_\SA$ and $\SD$ uses $Y_\SD$ resources at the NE. 
The partition of $(X_\SA, Y_\SD)$ enables us to understand when the attacker (resp. the defender) 
possesses sufficient amount of resources for the attack (resp. defence).

\vspace{-0.2cm}
\begin{table}[!htb]
\centering
\begin{tabular}{|c|c|}
\hline
$X_\SA^*$ & $:=\sum_{i=1}^{N}x_i^*$ (Tot\_Res used by $\SA$ at the NE)\\
\hline
$Y_\SD^*$ & $:=\sum_{i=1}^{N}y_i^*$ (Tot\_Res used by $\SD$ at the NE)\\
\hline $X_{\SA}^{suf}$ & Tot\_Res used by 
$\SA$ at the NE with $\lambda{=}\rho{=}0$\\
\hline $Y_{\SD}^{suf}$ & Tot\_Res used by 
$\SD$ at the NE with $\lambda{=}\rho{=}0$ \\
\hline $\hat{X}^{suf}_\SA$ & Tot\_Res needed 
by $\SA$ at the NE to let $\lambda{=}0$, \\
& given $Y_\SD < Y_{\SD}^{suf}$ (i.e. $\rho{>}0$)
\\
\hline $\hat{Y}^{suf}_\SA$ & Tot\_Res needed 
by $\SD$ at the NE to let $\rho{=}0$, \\
& given $X_\SA < X_{\SA}^{suf}$ 
(i.e. $\lambda{>}0$)
\\
\hline
\end{tabular}
\caption{Notations of total amount of resources}
\label{table:variables1}
\vspace{-0.3cm}
\end{table}
\vspace{-0.2cm}

The remaining challenge on deriving the NE is how $\lambda$ and $\rho$ are found for the 
given $K_\SA$ and $K_\SD$. Intuitively, we can solve $\lambda$ and $\rho$ based on Eqs. 
\eqref{eq:slack1}\eqref{eq:slack2}\eqref{eq:xy_ne1}\eqref{eq:xy_ne2}. However, there does 
not exist an explcit expression in general. We propose a bisection algorithm in 
Fig. \ref{figure:algorithm} to search $\lambda$ and $\rho$. The basic idea is to express $\rho$
as two functions of $\lambda$, $\rho_1(\lambda)$ obtained from Eqs. \eqref{eq:slack1}\eqref{eq:xy_ne1}\eqref{eq:xy_ne2}
and $\rho_2(\lambda)$ obtained from Eqs. \eqref{eq:slack2}\eqref{eq:xy_ne1}\eqref{eq:xy_ne2}, 
and then 
compute their intersection. To guarantee that the bisection 
algorithm can find feasible $\lambda$ and $\rho$ if they exist, we show the monotonicity 
of $\rho_1(\lambda)$ and $\rho_2(\lambda)$ in the following lemma.

\begin{lemma} 
Suppose that feasible $\lambda$ and $\rho$ (i.e. $\lambda,\rho{\geq} 0$) exist for the 
fixed $K_\SA$ and $K_\SD$ at the NE. The following properties hold 
i) if $\lambda$ is 0, there has a unique $\rho \geq0$; ii) if $\rho$ is 0, there has a unique $\lambda$; 
iii) $\rho_1(\lambda)$ is a strictly increasing function and $\rho_2(\lambda)$ is a  
strictly decreasing function.
\label{lemma:monotonicity}
\end{lemma}

The monotonicity property enables us to use bisection algorithm 
to check the existence of the pair $(\lambda, \rho)$
and solve them if they exist. When $X_\SA$ and $Y_\SD$ are sufficient, the NE
can be directly computed via eqs.\eqref{eq:xy_ne1} and \eqref{eq:xy_ne2}. 
When the resources of either 
$\SA$ or $\SD$ are insufficient, the NE is found 
by the lines $5{\sim} 17$ in Fig.\ref{figure:algorithm}. 
When both players have insufficient resources, the NE
is obtained by the lines $18{\sim}26$. The complexity
order of finding the sets with positive resource
allocation is merely $O(N^2)$. 

\vspace{-0.6cm}

\begin{figure}[htb]
\begin{tabbing}
\hskip 0.12in \=xx\=x\=x\=x\=x\=x\=x\=x\kill
\rule{3.3in}{0.25mm}\\
 {\bf Input: $N$, $X_\SA$, $Y_\SD$, $w_i$, $c$, $\hat{c}$, $f(\cdot)$ and $g(\cdot)$;} \\
 {\bf Output: $K_\SA$, $K_\SD$, $\lambda$, $\rho$, $x_i^*$ and $y_i^*$}\\
1: {\bf Initialize} $K_\SA = K_\SD = N$\\
2: Let $\lambda{=}\rho{=}0$, compute $y_i^{*}$, $x_i^{*}$ using eqs. \eqref{eq:xy_ne1},\eqref{eq:xy_ne2} for all $i$;\\
3: Compute $X_\SA^{suf}:=\sum_{i=1}^{N}x_i^*$ and $Y_\SD^{suf} = \sum_{i=1}^{N}y_i^*$;\\
4. \textbf{If} both $X_\SA\geq X_\SA^{suf}$ and $Y_\SD\geq Y_\SD^{suf}$, \textbf{exit};\\   
5: {\bf For} $K_\SA \geq 1$ \\
6: \> \> $K_\SD = K_\SA$ \\
7: \> \> {\bf For} $K_\SD \geq 1$ \\
8: \> \> \> \bf{If} $X_\SA \leq X_\SA^{suf}$\\
9: \>\>\>\> Find $\lambda$ by letting $\rho = 0$ and $X_\SA^* = X_\SA$ via \eqref{eq:xy_ne1}\eqref{eq:xy_ne2};\\
10:\>\>\> \bf{Elseif} $Y_\SD \leq Y_\SD^{suf}$\\
11: \>\>\>\> Find $\rho$ by letting $\lambda = 0$ and $Y_\SD^* = Y_\SD$ via \eqref{eq:xy_ne1}\eqref{eq:xy_ne2};\\
12: \>\>\> \bf{End};\\
13: \>\>\> \bf{If} $x_i^* \geq 0$, $y_i^* \geq 0$, \bf{exit};\\
14: \> \> \> $K_\SD = K_\SD - 1$\\
15:\> \> {\bf End}\\
16: \> \> $K_\SA = K_\SA - 1$\\
17: {\bf End}\\
18: {\bf For} $K_\SA \geq 1$ \\
19: \> \> $K_\SD = K_\SA = N$ \\
20: \> \> {\bf For} $K_\SD \geq 1$ \\
21: \> \> \> Compute the fixed point ($\rho, \lambda$) which solves \eqref{eq:xy_ne1} \\
\> \> \>  and \eqref{eq:xy_ne2} by setting $Y_\SD^* {=} Y_\SD$ and $X_\SA^* {=} X_\SA$; Given \\
\> \> \>  new pair $(\lambda,\rho)$, compute $y_i^*$ and $x_i^*$ via \eqref{eq:xy_ne1} and \eqref{eq:xy_ne2};\\
22: \> \> \bf{If} $x_i^* \geq 0$, $y_i^* \geq 0$, \bf{exit};\\
23: \> \> $K_\SD = K_\SD - 1$\\
24:\> \> {\bf End}\\
25: \> \> $K_\SA = K_\SA - 1$\\
26: {\bf End}\\
\rule{3.3in}{0.25mm}
\end{tabbing}
\vspace{-0.4cm}
\caption{Algorithm to find $K_\SA$, $K_\SD$, $\lambda$, $\rho$, $x_i^*$ and $y_i^*$ at the NE}
\label{figure:algorithm}
\vspace{-0.4cm}
\end{figure}

\subsection{Properties of NE}

Given the resource limits $X_\SA$, $Y_\SD$ and other system parameters,
we now know the way that the unique NE is computed. 
Our subsequent question is how a player 
disposes resources on heterogeneous targets at the NE. 

\begin{lemma} 
The NE $(\bx^*,\by^*)$ satisfies the following 
properties:
\begin{itemize}
\item $y_i^* \geq y_j^*$ for $1\leq i<j\leq K_\SD$;
\item $x_i^* \geq x_j^*$ for $K_\SD <i<j\leq K_\SA$;
\item i) $x_i^*>x_j^*$ if $\frac{\tilde{g}'(y)}{\tilde{g}(y)}$ is 
strictly increasing w.r.t. $y$, ii) $x_i^*=x_j^*$  
if $\frac{\tilde{g}'(y)}{\tilde{g}(y)}$ is a constant, and iii) 
$x_i^*<x_j^*$ if $\frac{\tilde{g}'(y)}{\tilde{g}(y)}$ is strictly 
decreasing w.r.t. $y$ for all $1{\leq}i{<}j{\leq} K_\SD$.
\end{itemize}
\label{lemma:neproperty}
\end{lemma}
The first property manifests that $\SD$ is inclined to allocate more resources 
to the targets with higher 
weights at the NE. The second property means 
that if two targets are not protected by $\SD$ at the 
NE, $\SA$ allocates more resources to the one of higher value. However, it is uncertain 
whether $\SA$ allocates more (or less) resources to a high (or lower) value target among 
the top $K_\SD$ targets with positive resources of $\SD$. 
We next use three examples to highlight that all the possibilities can happen. 
These examples differ in the choice of (complementary) defence efficiency functions. 
We define a new term, ``relative ineffectiveness of defence (RID)'', as the expression 
$\frac{\tilde{g}'(y)}{\tilde{g}(y)}$. Note that the first-order derivative $\tilde{g}'(y)$ reflects how fast (i.e. the slope)
$\tilde{g}(y)$ decreases with the increase of $y$. RID reflects the relative slope that the 
increase of $y$ reduces $\tilde{g}(y)$. If $\frac{\tilde{g}'(y)}{\tilde{g}(y)}$ is decreasing in $y$, further increasing $y$ 
makes $\tilde{g}(y)$ decreases faster and faster. On the contrary, if $\frac{\tilde{g}'(y)}{\tilde{g}(y)}$
is increasing in $y$, further increasing $y$ only results in a 
smaller and smaller relative reduction of $\tilde{g}(y)$ (considering the sign of $\tilde{g}'(y)$). 
We suppose that $\SA$ and $\SD$ allocate positive resources to $B_1$ and $B_2$. 

\noindent \textbf{Example 1 (InvG):} $f(x) {=} 1{-}(1{+}x)^{-a}$ and $\tilde{g}(y) {=} \frac{1}{1{+}\theta y}$. 
The following defence inefficiency 
equality holds,  $\frac{\tilde{g}'(y)}{\tilde{g}(y)} {=} \frac{-\theta}{1{+}\theta y}$. Then, we obtain 
$\frac{w_i}{w_j}=(\frac{1+x_i}{1+x_j})^{2(1{+}a)}\frac{1-(1{+}x_i)^{-a}}{1-(1{+}x_j)^{-a}}$. 
Due to $w_i>w_j$, it is easy to show $x_i>x_j$ by contradiction.

\noindent \textbf{Example 2 (ExpG):} $f(x) {=} 1{-}(1{+}x)^{{-}a}$ and 
$g(y) {=} \exp(-\theta y)$. The expression $\frac{\tilde{g}'(y)}{\tilde{g}(y)}$
is equal to $-\theta$. According to the KKT conditions 
in Theorem \ref{theorem:ne_condition1}, there has $(\frac{1{+}x_i}{1{+}x_j})^{1{+}a}\frac{1{-}(1+x_i)^{{-}a}}{1{-}(1+x_j)^{{-}a}}=1$.
The above equation holds only upon $x_i=x_j$.

\noindent \textbf{Example 3 (QuadG):} $f(x) {=} 1-(1+x)^{-a}$ and 
$\tilde{g}(y) = (1-\theta y)^2$. There exists $\frac{\tilde{g}'(y)}{\tilde{g}(y)} {=} -\frac{2\theta}{1-\theta y}$. Theorem \ref{theorem:ne_condition1}
yields $\frac{w_j}{w_i}=(\frac{1{+}x_i}{1{+}x_j})^{1{+}a}(\frac{1{-}(1+x_i)^{{-}a}}{1{-}(1+x_j)^{{-}a}})^2$.
Then, there has $x_i^*<x_j^*$.

\noindent \textbf{Remark 2:} For InvG-like $\tilde{g}(y)$, RID is strictly increasing. 
The attacker's best strategy is to allocate more resources to more important targets. 
In a word, the attacker and the defender have a ``head-on confrontation''. 
For ExpG-like $\tilde{g}(y)$, RID is a 
constant. The attacker sees a number of equally profitable targets. For QuadG-like $\tilde{g}(y)$, 
RID is a decreasing function. The attacker tries to avoid the targets that are \emph{effectively} 
protected by the defender.

Intuitively, when a player does not possess sufficient resources, he will gain a higher utility 
if his resource limit increases. This is true in a variety of cases. 
Suppose that not all the targets are attacked by $\SA$. When $X_\SA$ increases, $\SA$
can at least gain more profits by allocating the extra resources to the targets that are not under attack.
We next present a counter-intuitive example. 
Suppose that $\SA$ and $\SD$ allocate positive amount of resources to all the targets at the NE. 
The resources of $\SA$ are insufficient while those of $Y_\SD$ are sufficient, that is,
$\lambda>0$ and $\rho=0$. When $X_\SA$ increases, it is easy to show by contradiction 
that $\lambda$ decreases and 
$x_i$ increases. Due to the equality $-w_if(x_i)\tilde{g}'(y_i) = \hat{c}$ in the KKT conditions, 
$y_i$ also becomes larger. 
The utility of the attacker on target $B_i$ at the NE is given by $w_if(x_i)\tilde{g}(y_i)-cx_i = 
-\hat{c}\frac{\tilde{g}(y_i)}{\tilde{g}'(y_i)} - cx_i$. 
If RID of the defender, $\frac{\tilde{g}'(y)}{\tilde{g}(y)}$, is a constant or a decreasing function of $y_i$,
the expression $-\hat{c}\frac{\tilde{g}(y_i)}{\tilde{g}'(y_i)}$ is a constant or decreases as $y_i$ increases. Hence, the 
utility of the attacker on target $B_i$ decreases when $X_\SA$ increases. 
 
\noindent \textbf{Remark 3:} When the defender's resources are insufficient, the attacker gains 
more profits by acquiring more resources and allocating them to more important targets. When 
the defender's resources are sufficient, the attacker may explore new targets 
to attack, other than using all the resources to battle with the resource sufficient defender 
at the NE.

\subsection{Visualizing Whether a Target Is Attacked or Protected}

From Theorem \ref{theorem:ne_condition1}, one can 
see that $x_i^*$ and $y_i^*$ may take on 0, 
which occurs when $\SA$ or $\SD$ decides not to attack or defend target $B_i$.
We next show
the regions of $\lambda$ and $\rho$ upon which $x_i^*$ or $y_i^*$ hits 0. 
There are four possibilities,
i) $x_i^*=0$ and $y_i^* = 0$; ii) $x_i^*>0$ and $y_i^* = 0$; iii) $x_i^*=0$ and $y_i^* > 0$; and iv) $x_i^*>0$ and $y_i^*>0$.
We denote $R_{++}:=\{\lambda\geq 0; \rho\geq 0\}$. 

\noindent {\bf Case (i):} 
Eqs. \eqref{eq:theorem2_no1} and \eqref{eq:theorem2_no2} yield 
the region $R_1(\lambda,\rho)$, wherein
both $\SA$ and $\SD$ give up target $B_i$.

$
\quad\quad\quad\quad\quad R_1(\lambda,\rho) = \{\lambda \geq \max(w_if'(0)\tilde{g}(0) - c, 0) ; \rho \geq \max(-w_if(0)\tilde{g}'(0) - \hat{c}, 0)\}.
$

\noindent
{\bf Case (ii):} Given $x_i^* = h_\SA(\frac{c+\lambda}{w_i\tilde{g}(0)})$,
we obtain the region $R_2(\lambda,\rho)$ wherein $\SA$ attacks but
$\SD$ gives up target $B_i$:

$
\quad\quad\quad\quad\quad R_2(\lambda,\rho) = \{0\leq \lambda < w_if'(0)\tilde{g}(0) - c; \rho  > \max(-w_if\big(h_\SA(\frac{c+\lambda}{w_i\tilde{g}(0)})\big) \tilde{g}'(0) - \hat{c},0)    \};
$

\noindent
{\bf Case (iii):} Substituting 
$y_i^*$ by $h_\SD(-\frac{\rho{+}\hat{c}}{w_if(0)})$ 
in Eq. \eqref{eq:xy_ne2}, 
we obtain the region $R_3(\lambda,\rho)$ where $\SA$ gives up while
$\SD$ defends target $B_i$:

$
\quad\quad\quad\quad\quad R_3(\lambda,\rho) = \{0\leq \rho < -w_if(0)\tilde{g}'(0) - \hat{c}; \lambda  > \max(w_if'(0)\tilde{g}\big(h_\SD(-\frac{\rho+\hat{c}}{w_if(0)})), 0)\};
$

\noindent 
Due to $f(0):=0$, $\rho$ does not possess a valid value, so the region $R_3$ is empty.

\noindent 
{\bf Case (iv):} We have the region $R_4 = R_{++}\setminus \{R_1\cup R_2\cup R_3\}$.
For any $i=1,{\cdots}, N$, $x_i^*$ and $y_i^*$ contain two unknown variables $\lambda$ and $\rho$. Hence, in case iv), we can rewrite
$x_i^*$ and $y_i^*$ by $x_i^*(\lambda,\rho)$ and $y_i^*(\lambda,\rho)$.

\noindent
{\bf Remark 4:} The physical meanings of $R_1$ to $R_4$ are as follows: 
i) if $(\lambda,\rho)\in R_1$, both $\SA$ and $\SD$ do not 
     allocate resources to this target; 
ii) if $(\lambda,\rho)\in R_2$, $\SA$ attacks this target while
    $\SD$ decides not to defend it; 
iii) if $(\lambda,\rho)\in R_3$, $\SA$ does not attack this target 
     while $\SD$ defends it; 
iv) if $(\lambda,\rho)\in R_4$, $\SA$ attacks this target 
    and $\SD$ defends it. 

The purpose of defining $R_1$ to $R_4$ is that
we can gain some insights into the impacts of system parameters 
(e.g., $X_\SA$, $Y_\SD$, $w_i$, $c$ and $\hat{c}$)
on the NE without directly solving the NE. 
Here, for any pair $(\lambda,\rho)$, the increase of $\lambda$ means 
the decrease of $X_\SA$, and the increase of $\rho$ means the decrease 
of $Y_\SD$. This property is derived in the proof of Lemma \ref{lemma:monotonicity}. 
Let us illustrate $R_1\sim R_4$ by using a simple example.

\vspace{-0.5cm}
\begin{figure}[htb]
    \centering
   \includegraphics[width=2in, height = 1.6in]{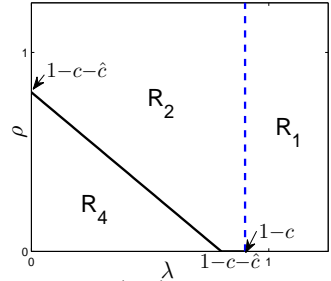}
   \vspace{-0.5cm}
   \caption{$(\lambda,\rho)$ regions of Ex.5}
   \label{fig:regions_example2}
\end{figure}

\noindent \textbf{Example 4:} $f(x) {=} 1{-}\exp(-x); \tilde{g}(y) {=} \exp(-y); w_i {=} 1.$ 
It is easy to obtain $h_\SA(x) {=} -\log(x)$ and $h_\SD(y) {=} {-}\log(-y).$
Substituting these expressions to 
$R_1$-$R_4$ in the above equations,
we derive the regions of $(\lambda,\rho)$ by
$
R_1(\lambda,\rho) = \{ \lambda \geq \max(1-c, 0), \rho \geq 0\}; R_3 = \emptyset
$
and 
$
R_2(\lambda,\rho) = \{0\leq \lambda < 1-c, 0\leq \rho < 1-c-\hat{c}-\lambda\}; R_4 =  R_{++}\setminus \{R_1\cup R_2\}.
$
Fig. \ref{fig:regions_example2} shows these regions with parameters $c$ and $\hat{c}$. 
$R_1$ is not empty. This implies that both $\SA$ and $\SD$ 
do not allocate resource to this target when $\lambda$ is larger than $1{-}c$.  
A large $\lambda$ corre   -sponds to the situation that $X_\SA$ is relatively 
small compared with the optimally needed resources for the attacks. 
The increase of $\lambda$ drives a point in $R_4$ to migrate to $R_2$ or even $R_1$. 
This means that the defender and the attacker may give up this target in sequence
when $X_\SA$ becomes more and more scarce.

\section{{\bf A Linear Intrusion Detection Game for Product-form Model}}
\label{sec:intrusion}

In this section, we investigate the existence and uniqueness of NE 
of an intrusion detection game where the attack and defence efficiencies are linear functions.

\subsection{A Matrix-form Game}

We study a matrix-form multi-dimensional intrusion detection game. 
The payoff matrix on target $B_i$ is shown in Fig.\ref{fig:payoffmatrix}
where $A$ (resp. $NA$) denotes ``attack'' (resp. ``not attack'') strategy, and $D$ (resp. $ND$)
denotes ``defend'' (resp. ``not defend'') strategy. 
Here, $w_i$ denotes the loss of victims for the pair-wise strategies $(A, ND)$ and $\gamma w_i$ denotes
that for $(A, D)$ with $\gamma\in (0, 1)$.  
Let $c$ and $\hat{c}$ be the costs of the ``attack'' and the ``defend'' strategies. Note that $\hat{c}$
refers to not only the cost of resources, but also the cost of performance such as QoS or false alarm
of benign events. 
We consider the mixed strategies of $\SA$ and $\SD$ in which 
$\SA$ attacks target $B_i$ with probability $x_i$ and $\SD$ detects this target with 
probability $y_i$. Each player only has one action on all the targets, which yields
the resource constraints: $\sum\nolimits_{i=1}^N x_i\leq X_{\SA} \leq 1$, $\sum\nolimits_{i=1}^N y_i  \leq Y_{\SD} \leq 1$
and $0\leq x_i,y_i\leq 1$.

To make the game non-trivial, we let $\gamma w_{i} \leq c$ and $w_i > c$ $\forall i$, i.e. 
the loss of victims is greater than the cost of the attacker on an unprotected target, and is 
less than this cost on a protected target.
Given the attack probabilities $\{x_i\}_{i=1}^N$ and the detection probabilities $\{y_i\}_{i=1}^N$, 
the utilities of $\SA$ and $\SD$ can be derived easily,
\begin{eqnarray}
U_{\SA} \!\!&=&\!\! w_{i}x_i - (1-\gamma)w_{i}x_iy_i - cx_i, \nonumber \\
U_{\SD} \!\!&=&\!\! -w_{i}x_i + (1-\gamma)w_{i}x_iy_i - \hat{c}y_i \nonumber.
\end{eqnarray}
The above utility functions fall in the category of our product-form game with 
$f(x) := x$ and $\tilde{g}(y):= 1 - (1-\gamma) y$. 
The resource constraints hold naturally because the sum of attack probabilities is no larger than 1,
and the sum of detection probabilities is also no larger than 1.
For the sake of simplicity, we denote a new variable as $\bar{\gamma}:=1-\gamma$.

\begin{figure}
\centering
\begin{tabular}{l l l}
     &$D$ & $ND$\\
    $A$ & $(\gamma w_{i}-c, -\gamma_1 w_{i} -\hat{c})$ & $(w_{i}-c, -w_{i})$ \\
    $NA$ &$(0, -\hat{c})$ & $(0,0)$ \\
\end{tabular}
\caption{Payoff Matrix}
\vspace{-0.7cm}
\label{fig:payoffmatrix}
\end{figure}

\subsection{Computing NE}

We take the derivatives of $U_\SA$ (resp. 
$U_\SD$) over $x_i$ (resp. $y_i$) and obtain
{
\setlength{\abovedisplayskip}{2pt}
\setlength{\belowdisplayskip}{2pt}
\begin{eqnarray}
dU_\SA/dx_i \!\!&=&\!\! w_i - w_i\bar{\gamma}y_i -c, \nonumber\\
dU_\SD/dy_i \!\!&=&\!\! w_i\bar{\gamma}x_i -\hat{c}.
\nonumber
\label{eq:linear1}
\end{eqnarray}
}

The existence of a NE is guaranteed by the concavity of the game. 
Before 
diving into the solution of the NE, we present a property of the sets of targets 
that are attacked or defended at the NE.

\begin{lemma} 
The sets of targets with positive resources at the NE are given by
i) $\{B_1,\cdots, B_{K_\SA}\}$ for the attacker and $\{B_1,\cdots, B_{K_\SD}\}$ 
for the defender; 
ii) either $K_\SA = K_\SD$ or $K_\SA = K_\SD+1$.
\label{lemma:linear_neproperty1}
\end{lemma}

Lemma \ref{lemma:linear_neproperty1} is the sufficient 
condition of the existence of NE. Similar to 
Lemma \ref{lemma:attackdefendsets}, 
$\SA$ and $\SD$ allocate
resources to the subsets of more important targets. 
The difference lies in that $\SA$ may allocate resources to more targets than $\SD$ when 
$f(\cdot)$ and $g(\cdot)$ are nonlinear functions, but 
to at most one more target than $\SD$ when 
$f(\cdot)$ and $g(\cdot)$ are our linear functions. We proceed 
to find the NE by considering different regions of 
$X_\SA$ and $Y_\SD$ in the following theorem.

\begin{theorem}
The multi-dimensional intrusion detection game admits a NE as below
\begin{itemize}
\item  $P_\SA(k) \!{<}X_\SA \!{<}\! P_\SA(k{+}1)$ and $Y_\SD \!{>} P_\SD(k{+}1)$ for $0{\leq}\!k{\leq}\!N{-}1$.
The NE is uniquely determined by
{
\setlength{\abovedisplayskip}{5pt}
\setlength{\belowdisplayskip}{5pt}
\begin{eqnarray}
\!\!\!\!\!x_i^*\!\!\!&=&\!\!\!\left\{\begin{matrix}
\;\frac{\hat{c}}{w_i\bar{\gamma}}, & \forall \; i\leq k \\
\;X_\SA{-}\sum_{j=1}^k \frac{\hat{c}}{w_j}, & i {=} k{+}1 \\
\;0, & \forall \; i {>} k{+}1 
\end{matrix}\right. 
\label{eq:theorem3_no1.1}\\
\!\!\!\!\!y_i^* \!\!\!&=&\!\!\! \left\{\begin{matrix}
\; (1-\frac{w_{k{+}1}}{w_i})\frac{1}{\bar{\gamma}} , & \forall \; i{\leq} k \\
\;0, & \forall \; i > k 
\end{matrix}\right. .
\label{eq:theorem3_no1.2}
\end{eqnarray}
}
\noindent Here, the sum over an empty set is 0 conventionally.
\item  $P_\SD(k) {<} Y_\SD {<} P_\SD(k{+}1)$ and $X_\SA {>} P_\SA(k)$ for $1{\leq}k{\leq}N$.
The NE is uniquely determined by
{
\setlength{\abovedisplayskip}{5pt}
\setlength{\belowdisplayskip}{5pt}
\begin{eqnarray}
\!\!\!\!\!x_i^*\!\!&=&\!\!\left\{\begin{matrix}
\;(\sum_{j=1}^{k}\frac{w_i}{w_j})^{-1}X_\SA, & \forall \; i\leq k \\
\;0, & \forall \; i > k
\end{matrix}\right. 
\label{eq:theorem3_no2.1}\\
\!\!\!\!\!y_i^* \!\!&=&\!\! \left\{\begin{matrix}
 (\sum\nolimits_{j=1}^{k}\frac{w_i}{w_j})^{-1}
\big(Y_\SD  {-} \frac{1}{\bar{\gamma}}k \big) {+} \frac{1}{\bar{\gamma}} , & \!\!\forall  i{\leq} k \\
0, & \!\!\forall  i {>} k 
\end{matrix}\right. .
\label{eq:theorem3_no2.2}
\end{eqnarray}
}
\item $X_\SA {>} P_\SA(N)$ and $Y_\SD {>} P_\SD(N{+}1)$.

\noindent The NE is uniquely determined by 
{
\setlength{\abovedisplayskip}{5pt}
\setlength{\belowdisplayskip}{5pt}
\begin{eqnarray}
x_i^*= \frac{\hat{c}}{w_i\bar{\gamma}},\;\; y_i^* = \frac{1}{\bar{\gamma}} {-} \frac{c}{w_i\bar{\gamma}},\;\; \forall \;\; 1{\leq}i{\leq}N .
\label{eq:theorem3_no3}
\end{eqnarray}
}
\item $X_\SA {=} P_\SA(k)$ and $Y_\SD {\geq} P_\SD(k)$ for $1{\leq}k{\leq}N$. 

\noindent Denote by $\tilde{Y}_\SD$ an arbitrary real number 
in the range $[P_\SD(k), \min\{Y_\SD, P_\SD(k{+}1) \}]$. A NE is given by
{
\setlength{\abovedisplayskip}{5pt}
\setlength{\belowdisplayskip}{5pt}
\begin{eqnarray}
\!\!\!\!\!\!x_i^*\!\!\!\!&=&\!\!\!\!\left\{\begin{matrix}
\;\frac{\hat{c}}{w_i\bar{\gamma}}, & \forall \;\; i\leq k \\
\;0, & \forall \;\;  k{+}1{\leq} i{\leq}N
\end{matrix}\right.  \label{eq:linearne7}
\\
\!\!\!\!\!\!y_i^* \!\!\!\!&=&\!\!\!\! \left\{\begin{matrix}
\; \frac{1}{\bar{\gamma}} {+} (\sum\nolimits_{j{=}1}^k \frac{w_i}{w_j})^{-1}(\tilde{Y}_\SD {-} k\frac{1}{\bar{\gamma}}), & \forall \; i\leq k \\
\;0, & \forall \;i > k 
\end{matrix}\right. . \label{eq:linearne8}
\end{eqnarray}
}
\item $Y_\SD {=} P_\SD(k)$ and $P_\SA(k{-}1) {\leq} X_\SA {\leq} P_\SA(k)$ for $2{\leq}k{\leq}N$. 

\noindent Denote by $\tilde{X}_\SA$ an arbitrary real number 
in the range $[P_\SA(k{-}1), X_\SA]$. A NE is given by
{
\setlength{\abovedisplayskip}{5pt}
\setlength{\belowdisplayskip}{5pt}
\begin{eqnarray}
x_i^*\!\!\!&=&\!\!\!\left\{\begin{matrix}
\;(\sum\nolimits_{j{=}1}^k \frac{w_i}{w_j})^{-1}\tilde{X}_\SA, & \forall \; i\leq k \\
\;0, & \forall \; i {>} k{+}1 
\end{matrix}\right.  \label{eq:linearne9}
\\
y_i^* \!\!\!&=&\!\!\! \left\{\begin{matrix}
\; (1{-}\frac{w_{k{+}1}}{w_i})\frac{1}{\bar{\gamma}} , & \forall \; i\leq k \\
\;0, & \forall \; i > k 
\end{matrix}\right..  \label{eq:linearne10}
\end{eqnarray}
}
\end{itemize}
Here, $P_\SA(k)$ and $P_\SD(k)$ are defined as 
$P_\SA(0) {:=}0$, 
$P_\SA(k){:=}\sum\nolimits_{i{=}1}^k\frac{\hat{c}}{w_i\bar{\gamma}}, \; \forall \; 1{\leq}k{\leq}N$; $P_\SD(1) {=} 0$,  $P_\SD(k){:=}$

\noindent $\sum\nolimits_{i{=}1}^{k{-}1}\frac{1}{\bar{\gamma}}(1{-}\frac{w_{k}}{w_i})$, and $P_\SD(N{+}1)  {:=} \frac{1}{\bar{\gamma}}N{-}\sum\nolimits_{i{=}1}^{N}\frac{c}{w_i\bar{\gamma}}$. 
\label{theorem:ne_linear}
\end{theorem}

\begin{figure}[htb]
\vspace{-0.6cm}
    \centering
   \includegraphics[width=2.0in]{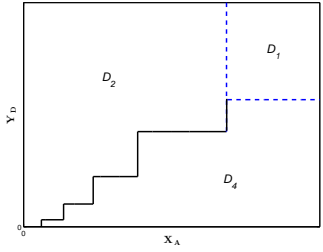}
   \vspace{-0.3cm}
   \caption{Sufficiency of $X_\SA$ and $Y_\SD$ with linear $f(\cdot)$ and $g(\cdot)$}
   \label{fig:linear_zones}
   \vspace{-0.3cm}
\end{figure}

We illustrate the relationship between NE and resource 
limits in Fig.\ref{fig:linear_zones}.  When $f(\cdot)$ and $g(\cdot)$
are linear, the best response of a player becomes a step-like function. 
The feasible domain of $(X_\SA,Y_\SD)$ is partitioned into three parts: 
i) $D_1$ {-} sufficient $X_\SA$ and sufficient $Y_\SD$; ii) $D_2$ - 
insufficient $X_\SA$ and sufficient $Y_\SD$; 
iii) $D_4$ - insufficient $X_\SA$ and insufficient $Y_\SD$. 
The total consumed resources at the NEs for $D_1$ and 
$D_2$ are located in the step-like boundary curve. 
When $X_\SA$ or $Y_\SD$ take some special values, 
the boundary curve illustrates the 
existence of multiple NEs. In the horizontal boundary,  different NEs bring the 
same utility to the attacker, but different utilities to the defender. In the vertical 
boundary, the utilities of the defender are the same, while those of the attacker
are different. 
Let us take a look at an example with $X_\SA{=}\frac{\hat{c}}{w_1\bar{\gamma}}$ and 
$Y_\SD{>}(1{-}\frac{w_2}{w_1})\frac{1}{\bar{\gamma}}$. Two NEs on target $B_1$ 
can be $(x_1^*, y_1^*)_{(1)} {=} (\frac{\hat{c}}{w_1\bar{\gamma}}, 0)$ and $(x_1^*,y_1^*)_{(2)} {=} 
(\frac{\hat{c}}{w_1\bar{\gamma}},(1{-}\frac{w_2}{w_1})\frac{1}{\bar{\gamma}})$. Both $\SA$ and $\SD$
do not allocate resources to other targets. 
The utility of $\SD$ is given by $U_\SD {=} {-}\frac{1}{\bar{\gamma}}\hat{c}$ at the both NEs. 
The utilities of $\SA$ are given by $U_\SA^{(1)} {=} x_1^*(w_1{-}c)$ and 
$U_\SA^{(2)}{=}x_1^*(w_2{-}c)$ at the two NEs. At the first NE, 
$B_1$ is the most profitable to $\SA$. At the second NE, $B_1$ and $B_2$
are equally profitable. In both NEs, $\SA$ cannot gain more profits by switching to another
allocation strategy unilaterally. Besides, the total consumed resources for $D_4$ can be mapped to an arbitrary point in this domain, in which both players have insufficient resources.

\noindent \textbf{Remark 5:} We summarize the salient properties of the NEs for linear 
attacking efficiency and linear uptime as below. 

\noindent 1) The targets with $x_i^*>0$ are equally profitable to $\SA$ such 
that $\SA$ has no incentive to change his strategy. 

\noindent 2) $\SD$ prefers to allocate more resources to the more valuable targets. 
As a countermeasure, $\SA$ allocates more resources to the targets that are not 
effectively protected by $\SD$. 

\noindent 3) The NE is not unique with some special choices for $X_\SA$ and $Y_\SD$. 
If multiple NEs exist for a given pair $(X_\SA,Y_\SD)$, they yield the same utility for one player, but
different utilities for the other player.

\section{{\bf Nash Equilibrium for Proportion-form Model}}
\label{sec:coupled}

In this section, we analyze the NE strategy of the players 
on different targets for the proportion-form
breaching model.

\noindent \textbf{Nash Equilibrium and its Properties:}

We define $(\bx^*, \by^* )$ as the NE of the game for the proportion-form model. 
Here, we relax the constraints to be $f(\cdot),g(\cdot) {\geq} 0$  
(unlike $0{\leq} f(\cdot),g(\cdot){\leq} 1$ in the product-form model).
The breaching probability in the proportion-form model cannot exceed 1. 
Based on the KKT conditions, $(\bx^*, \by^* )$ is given by the following theorem.

\begin{theorem}
There exist non-negative variables $\lambda$ and $\rho$ such that
{
\setlength{\abovedisplayskip}{3pt}
\setlength{\belowdisplayskip}{3pt}
\begin{eqnarray}
w_i\frac{f(x_i^*)g'(y_i^*)}{(f(x_i^*)+g(y_i^*))^2}- \hat{c}  \;\; \left\{\begin{matrix}
\;=\rho \;\; &&\textrm{ if } \;\; y_i^* > 0 \\
\;\leq \rho \;\; &&\textrm{ if } \;\; y_i^* = 0 
\end{matrix}\right., 
\label{eq:theorem_coupled1}
\end{eqnarray}
\begin{eqnarray}
w_i\frac{f'(x_i^*)g(y_i^*)}{(f(x_i^*)+g(y_i^*))^2}  - c \;\; \left\{\begin{matrix}
\;=\lambda \;\; &&\textrm{ if } \;\; x_i^* > 0 \\
\;\leq \lambda \;\; &&\textrm{ if } \;\; x_i^* = 0 
\end{matrix}\right.,
\label{eq:theorem_coupled2}
\end{eqnarray}
}
\noindent with the slackness conditions in Eq.\eqref{eq:slack1} and \eqref{eq:slack2}.

\label{theorem:ne_coupled}
\end{theorem}

As the first step to find the NE, we need to investigate how many targets will be attacked by $\SA$ and
defended by $\SD$. The following lemma shows that both $\SA$ and $\SD$ allocate resources to 
all the targets in $\mathcal{B}$.
\begin{lemma}
At the NE, there have $x_i^*>0$ and $y_i^*>0$ for all $i=1,\cdots,K$ if $f(\cdot)$ and $g(\cdot)$ are concave and strictly 
increasing with $f(0)=0$ and $g(0) = 0$.
\label{lemma:coupled_targets}
\end{lemma}

Lemma \ref{lemma:coupled_targets} simplifies the complexity to obtain the NE strategy because 
we do not need to test whether a target will be attacked or defended. Then, the equalities in 
Eqs.\eqref{eq:theorem_coupled1} and \eqref{eq:theorem_coupled2} hold.
Similarly, we partition $(X_\SA, Y_\SD)$ into four domains to fine the NE: $\mathbf{D_1}$)
$X_\SA {\geq} X^{suf}_\SA$ and $Y_\SD {\geq} Y^{suf}_\SD$; 
$\mathbf{D_2}$) $X_\SA {<} X^{suf}_\SA$ and $Y_\SD {\geq} \hat{Y}^{suf}_\SD$;
$\mathbf{D_3}$) $X_\SA {\geq} \hat{X}^{suf}_\SA$ and $Y_\SD {<} Y^{suf}_\SD$; $\mathbf{D_4}$) none of the above. 
The method to find the NE contains the similar steps as those of the algorithm in Fig.\ref{figure:algorithm}.
We need to check whether $(X_\SA,Y_\SD)$ is located in a domain from $\mathbf{D_1}$ to $\mathbf{D_4}$ one by one.

We next study how $\SA$ and $\SD$ allocate resources to different targets, 
given the resource limits $X_{\SA}$ and $Y_{\SD}$. The NE strategy satisfies the following 
properties.
\begin{lemma}
$\SA$ and $\SD$ always allocate more resources to the more important targets, i.e. $x_i^* {>} x_j^*$ and $y_i^*{>}y_j^*$ 
if $w_i{>}w_j$.
\label{lemma:coupled_strategy}
\end{lemma}

\noindent \textbf{Remark 6:} In comparison to the product-form breaching model, 
the players in the proportion-form breaching model always allocate more resources to the more valuable targets.

For the generalized proportion-form breaching model, it is usually difficult to analyze 
how the NE and the utilities at the NE are influenced by the resource limits. Therefore, we consider 
two specific examples with explicit functions $f(\cdot)$ and $g(\cdot)$.

\noindent \textbf{Example 5:} Let $f(x)=x^a$ and $g(y)=y^a$ in the breaching probability model with $0{<}a{\leq} 1$. 
Then, for the four cases w.r.t. the sufficiency of $X_\SA$ and $Y_\SD$, there have:
\textbf{$\mathbf{D_1}$}): $X_\SA \geq X_\SA^{suf} = \frac{\sum_{i=1}^{N}w_ia(\frac{c}{\hat{c}})^a}{(1+(\frac{c}{\hat{c}})^a)^2\cdot c}$ 
and $Y_\SD \geq Y_\SA^{suf} =\frac{\sum_{i=1}^{N}w_ia(\frac{c}{\hat{c}})^a}{(1+(\frac{c}{\hat{c}})^a)^2\cdot \hat{c}} $. 
The increase of $X_\SA$ or $Y_\SD$ does not influence the NE and the utilities of $\SA$ and $\SD$. 
\textbf{$\mathbf{D_2}$)}: $X_\SA <X_\SA^{suf}$ and 
$Y_\SD \geq \hat{Y}_\SD^{suf} = \frac{\sum_{i=1}^Nw_ia(\frac{c{+}\lambda}{\hat{c}})^a}{(1+(\frac{c{+}\lambda}{\hat{c}})^a)^2\cdot \hat{c}}$. 
where $\lambda$ is determined by 
$$
(1+(\frac{c{+}\lambda}{\hat{c}})^a)^2\cdot \hat{c}^{a}(c+\lambda)^{1{-}a}X_\SA = \sum\nolimits_{i=1}^{N}w_i a.
$$
Due to $0{<}a{\leq}1$, $\lambda$ is a strictly decreasing function of $X_\SA$. As $X_\SA$ grows, $x_i^*$ and $y_i^*$ increase 
accordingly. 
Then, the utilities of $\SD$ and $\SA$ are given by
\begin{eqnarray}
U_\SD \!\!\!\!&=&\!\!\!\! -\sum\nolimits_{i=1}^N (\frac{w_i\hat{c}^a}{(c+\lambda)^a + \hat{c}^a} -  \hat{c}y_i^*); \nonumber\\
U_\SA \!\!\!\!&=&\!\!\!\! \sum\nolimits_{i=1}^N (\frac{w_i\hat{c}^a}{(c{+}\lambda)^a {+} \hat{c}^a} - 
\frac{w_iac(c{+}\lambda)^{(a-1)}\hat{c}^a}{((c{+}\lambda)^a {+} \hat{c}^a)^2}). \nonumber
\end{eqnarray}
It is obvious to see that $U_\SD$ is a decreasing function of the attacker's resource $X_\SA$. We take the first-order derivative 
of $U_\SA$ over $\lambda$. However, $U_\SA$ does not necessarily increase when $X_\SA$ grows. Let us take a 
look at a special situation with $a{=}1$. We then take the first order derivative of $U_\SA$ over $\lambda$ and obtain 
\begin{eqnarray}
\frac{dU_\SA}{d\lambda} \!\!\!\!&=&\!\!\!\! \sum\nolimits_{i=1}^N \frac{w_i\hat{c}(c{-}\lambda{-}\hat{c})}{(c{+}\lambda+\hat{c})^3}.
\end{eqnarray}
When $c{<}\lambda{+}\hat{c}$, $U_\SA$ is a decreasing function of $\lambda$, and hence an increasing function of $X_\SA$.
Otherwise, $U_\SA$ decreases as $X_\SA$ increases. This implies that $\SA$ always benefits from obtaining more resources 
if his cost is smaller than that of $\SD$. When $\SA$'s cost is larger than $\SD$'s, more resources may lead to a reduced utility of $\SA$.
\textbf{$\mathbf{D_3}$)}: $X_\SA > \hat{X}_\SA^{suf}$ and 
$Y_\SD \geq Y_\SD^{suf}$. This case is symmetric to that of $\mathbf{D_2}$), which is not analyzed here.
$\mathbf{D_4}$): both $X_\SA$ and $Y_\SD$ are insufficient. In this domain, all the resources of $\SA$ and $\SD$ are utilized. 
Then, there have $x_i^* = \frac{w_i}{\sum_{j=1}^Nw_j}X_\SA$ and $y_i^* = \frac{w_i}{\sum_{j=1}^Nw_j}Y_\SD$. 
The utilities of $\SA$ and $\SD$ are given by
\begin{eqnarray}
U_\SD \!\!\!\!&=&\!\!\!\! -\sum\nolimits_{i=1}^N \frac{w_i(X_\SA)^a}{(X_\SA)^a + (Y_\SD)^a}  -  \hat{c}Y_\SD; \nonumber\\
U_\SA \!\!\!\!&=&\!\!\!\! \sum\nolimits_{i=1}^N \frac{w_i(X_\SA)^a}{(X_\SA)^a + (Y_\SD)^a}  - cX_\SA
\end{eqnarray}
When $X_\SA$ increases, $U_\SD$ decreases accordingly. However, increasing $X_\SA$ does not necessarily bring a higher 
utility to $\SA$. Similarly, increasing $Y_\SD$ yields a worse utility to $\SA$, but not necessarily resulting a higher utility to $\SD$.

\section{{\bf Related Work}}
\label{sec:related}

Today's network attacks have evolved into online crimes such as phishing and mobile 
malware attacks. The attackers are profit-driven by stealing private information or even the 
money of victims. Authors in \cite{Moore1} measured
the uptime of malicious websites in phishing attacks to quantify the loss of victims. 
Sheng et al. provided the interviews of experts in \cite{Sheng09} to combat the phishing. A number of studies 
proposed improved algorithms to filter the spams containing links to malicious websites in \cite{Soldo10,Marchal12}.
In mobile platforms, users usually publish root exploits that can be leveraged by malicious attackers.
Authors in \cite{john1} proposed a new cloud-based mobile botnets to exploit push notifcation services as a means of command dissemination. 
They developed a stress test system to evaluate 
the effectiveness of the defence mechanisms for Android platform in \cite{john2}.
Felt et al. surveyed the behavior of current mobile malware and evaluated the effectiveness of 
existing defence mechanism in \cite{felt11}. 

Game theoretic studies of network security provide the fundamental understandings of
the decision making of attackers and defenders. Authors in \cite{zhu2} used stochastic
game to study the intrusion detection of networks. More related works on the network security game
with incomplete information and stochastic environment can be found in \cite{alpcan,survey2}.
Another string of works studied the security investment of nodes whose security level depended on the his security adoption and that of other nodes connected to him. Some models did not consider
the network topology \cite{chuang1} and some others studied either fixed
graph topologies \cite{omic} or the Poisson random graph \cite{john3,lelarge2}.

Among the studies of network security game, \cite{chuang3,tifs09,eitan1} are closely related to our work.
In \cite{chuang3}, authors used the standard Colonel Blotto game to study
the resource allocation for phishing attacks. An attacker wins a malicious website
if he allocates more resources than the defender, and loses otherwise. 
This may oversimplify the competition between an attacker and a defender. Our work differs 
in that the attackers perform attacks on multiple
non-identical banks or e-commerce companies, and the competition is modeled as a non-zero sum game that yields a pure strategy. 
In \cite{tifs09}, the authors formulated a linearized model for deciding the attack and monitoring 
probabilities on multiple servers in network intrusion attacks. 
Altman et al. in \cite{eitan1} studied a different type of 
multi-battlefield competition in wireless jamming attack that provides important insights of power allocation on OFDM channels.

\section{{\bf Conclusion}}
\label{sec:conclusion}

In this work, we formulate a generalized game framework to capture the conflict 
on multiple targets between a defender and an attacker that are resource constrained. 
A product-form and a proportion-form security breaching models are considered. 
We prove the existence of a unique NE, and propose efficient algorithms to search this NE 
when the game is strictly concave. 
Our analysis provides important insights in 
the practice of network attack and defence. For the \textit{product-form} breaching model,
i) the defender always allocates more resources to the more important target, while the attacker
may not follow this rule; ii) when the defender has sufficient amount of resources,  
more resources of the attacker might not bring a better utility to him;
iii) when the game is not strictly concave, 
there may exist multiple NEs that yield different utilities of the players.
For the \textit{proportion-form} breaching model, iv) both the attacker and the defender 
allocate more resources to more important targets; v) a resource insufficient 
player causes a reduction of his opponent's utility, while not necessarily gaining a 
better utility by himself when his resource limit increases.

\ifCLASSOPTIONcaptionsoff
  \newpage
\fi



%


\newpage

\twocolumn[  
    \begin{@twocolumnfalse}  
    \maketitle 
\section*{\Huge Supplement: Proofs of Lemmas and Theorems\\
{\Large Yuedong Xu, John C.S. Lui}\\{\tiny .}}
\label{sec:appendix1}
    \end{@twocolumnfalse}  
] 

\subsection*{Proof of Theorem \ref{lemma:existence}}

\noindent \textbf{Proof:} We prove the existence and uniqueness of the NE
for the product-form and the proportion-form breaching models separately. 

\noindent{\textit{Product-form:}} The second-order derivatives of $U_\SA$ 
over $\bx$ can be expressed as
\begin{eqnarray}
\frac{\partial^2U_\SA}{\partial x_i^2} = w_i f''(x_i)\tilde{g}(y_i) <0, 
\textrm{ and } 
\frac{\partial^2U_\SA}{\partial x_i\partial x_j} = 0, 
\;\;  \forall i,j. \nonumber
\end{eqnarray}
The second-order derivatives of $U_\SD$ 
over $\by$ are given by
\begin{eqnarray}
\frac{\partial^2U_\SD}{\partial y_i^2} = -w_i f(x_i)\tilde{g}''(y_i)<0, 
\textrm{ and } 
\frac{\partial^2U_\SD}{\partial y_i\partial y_j} = 0, 
\;\;  \forall i,j. \nonumber
\end{eqnarray}
Since $f(x_i)$ is strictly concave w.r.t. $x_i$, $U_\SA$ 
is a concave function of the strategy profile $\{x_i, i=1,\cdots, N\}$.
Based on Rosen's theorem \cite{rosen},
there always exists a NE in the game $\textbf{G}$. 

The matrix $M$ on a target is given by 
\begin{eqnarray}
M = w_i \begin{bmatrix}
       \varphi_1  f''(x)\tilde{g}(y) & \varphi_1 f'(x) \tilde{g}'(y)           \\
       -\varphi_2 f'(x)\tilde{g}'(y) & -\varphi_2 f(x)\tilde{g}''(y)    
     \end{bmatrix}.
\end{eqnarray}
Then, there has $M {+}M^T= $
\begin{eqnarray}
w_i \begin{bmatrix}
       2\varphi_1  f''(x)\tilde{g}(y) & (\varphi_1{-}\varphi_2) f'(x) \tilde{g}'(y)           \\
       (\varphi_1{-}\varphi_2) f'(x)\tilde{g}'(y) & -2\varphi_2 f(x)\tilde{g}''(y)    
     \end{bmatrix}.
\end{eqnarray}
Suppose $\varphi_1=\varphi_2 >0$. Because $f''(x) < 0$ and $\tilde{g}''(y)>0$, 
then matrix $-(M{+}M^T)$ is positive definite. Hence, $M{+}M^T$ is negative definite,
resulting in the unique NE in the game \textbf{G}.

\noindent\textit{Proportion-form:} 
The second-order derivatives of $U_\SA$ 
over $\bx$ can be expressed as
\begin{eqnarray}
\frac{\partial^2U_\SA}{\partial x_i^2} = w_ig(y_i)\frac{f''(x_i)(f(x_i){+}g(y_i)){-}2(f'(x_i))^2}{(f(x_i)+g(y_i))^3}<0 \nonumber
\end{eqnarray}
and $\frac{\partial^2U_\SA}{\partial x_i\partial x_j} = 0, 
\quad  \forall i,j$, due to $f''(x_i) < 0$.
The second-order derivatives of $U_\SD$ 
over $\by$ can be expressed as
\begin{eqnarray}
\frac{\partial^2U_\SD}{\partial y_i^2} = w_if(x_i)\frac{g''(x_i)(f(x_i){+}g(y_i)){-}2(g'(x_i))^2}{(f(x_i)+g(y_i))^3}<0 \nonumber
\end{eqnarray}
and $\frac{\partial^2U_\SD}{\partial y_i\partial y_j} = 0, 
\quad  \forall i,j,$ due to $g''(y_i)<0$. 
Hence, \textbf{G} is a concave game that admits a NE.

The matrix $M$ on a target is given by 
\begin{eqnarray}
M = \begin{bmatrix}
       \varphi_1 \frac{\partial^2U_\SA}{\partial x_i^2} & \varphi_1 \frac{\partial^2U_\SA}{\partial x_i\partial y_i}        \\
       \varphi_2 \frac{\partial^2U_\SD}{\partial x_i \partial y_i} &  \varphi_2 \frac{\partial^2U_\SD}{\partial y_i^2}  
     \end{bmatrix}.
\end{eqnarray}
Because of $\frac{\partial^2U_\SA}{\partial x_i\partial y_i}  = -\frac{\partial^2U_\SD}{\partial x_i \partial y_i}$, 
if we let $\varphi_1=\varphi_2>0$, the expression $M {+}M^T$ is obtained by
\begin{eqnarray}
M {+}M^T = \begin{bmatrix}
      2 \varphi_1 \frac{\partial^2U_\SA}{\partial x_i^2} & 0      \\
       0 &  2\varphi_2 \frac{\partial^2U_\SD}{\partial y_i^2}  
     \end{bmatrix}.
\end{eqnarray} 
It is obvious to see that $M {+}M^T$ is negative definite. Hence, in the proportion-form
breaching model, there exists a unique NE.  \done

\subsection*{Proof of Theorem \ref{theorem:ne_condition1}}

\noindent \textbf{Proof:} Recall that $U_\SA$ is concave in $\bx$
and $U_\SD$ is concave in $\by$. Then, the best responses
of $\SA$ and $\SD$ are the solutions to two
convex optimization problems. Let $\lambda$ and $\rho$ be Lagrange multipliers
of $\SA$ and $\SD$ respectively.
Let $L_\SD(\by,\rho)$ be the Lagrangian function of the defender $\SD$. 
We have
\begin{eqnarray}
L_\SD(\by,\rho) = -\sum\nolimits_{i=1}^{N} w_i f(x_i)\tilde{g}(y_i) - \hat{c} \sum\nolimits_{i=1}^{N} y_i \nonumber\\
+ \rho(Y_\SD - \sum\nolimits_{i=1}^N y_i).
\end{eqnarray}
Our first step is to find the optimal $y_j$ as a function of $\rho$.
Taking the derivative over $y_i$, we obtain
\begin{eqnarray}
\frac{dL_\SD(\by,\rho)}{dy_i} = {-}w_if(x_i)\tilde{g}'(y_i) {-} (\hat{c}{+}\rho) ,\;\;\forall i{=}1,{\cdots}, N.
\end{eqnarray}
The optimal resource allocated to target $i$, or $y_i$, satisfies the following condition
\begin{eqnarray}
\rho = -w_if(x_i)\tilde{g}'(y_i) - \hat{c}
\label{eq:computene1}
\end{eqnarray}
when $y_i$ is greater than 0. 
If $y_i=0$ and $\frac{dL_d(\textbf{y},\rho)}{dy_i} > 0$, we have:
\begin{eqnarray}
\rho > -w_if(x_i)\tilde{g}'(y_i) - \hat{c}.
\label{eq:eq:computene2}
\end{eqnarray}
When $\hat{c}$ is 0, the left hand of 
Equation~\eqref{eq:theorem2_no1} is positive for any $x_i > 0$.
Hence, $\rho$ is always positive if there is at least one target with $x_i>0$ at the NE.
This means that $\SD$ will consume all the resources $Y_\SD$.
When $\hat{c}>0$, the Karush-Kuhn-Tucker (KKT) conditions give rise to
\begin{eqnarray}
\left\{\begin{matrix}
\rho\geq 0 \;\; &&\textrm{ if } \;\; \sum\nolimits_{i=1}^N y_i^* = Y_\SD,\\
\rho = 0 \;\; &&\textrm{ if } \;\; \sum\nolimits_{i=1}^N y_i^* < Y_\SD.
\end{matrix}\right.
\end{eqnarray}

\noindent Following the same approach, we define the Lagrangian function of 
$\SA$ as
\begin{eqnarray}
L_\SA(\textbf{x},\lambda) =  \sum\nolimits_{i=1}^N w_i f(x_i) \tilde{g}(y_i) - c\sum\nolimits_{i=1}^N x_i \nonumber\\
+ \lambda (X_\SA {-} \sum\nolimits_{i=1}^N x_i).
\end{eqnarray}
The first-order derivatives are given by
\begin{eqnarray}
\frac{dL_\SA(\textbf{x},\nu)}{dx_i} = w_i f'(x_i)\tilde{g}(y_i) - (c+\lambda), \;\;\; \forall i = 1,\cdots, N.
\end{eqnarray}
If $x_i^*$ is non-zero, the above derivative equals to 0. Otherwise, 
$L_\SA(\textbf{x},\lambda)$ is a strictly decreasing
function of $x_i$ such that $x_i^*:=0$. The Lagrange multiplier $\lambda$ also satisfies the slackness condition. \done

\subsection*{Proof of Lemma \ref{lemma:attackdefendsets}}

\noindent \textbf{Proof:} Consider two targets $i,j$ with $w_i > w_j$. We assume that $x_i^*=0$ 
and $x_j^*>0$ at the NE. 
The utility received by $\SA$ is better if 
it shifts some $x_j^*$ to the $i^{th}$ target. 
This contradicts the assumption that the game is at the NE.
Hence, $\SA$ only attacks $K_\SA$ targets with 
the descending order of their weights. 

We next assume $y_i^*=0$ and $y_j^*>0$. There exists an inequality $-\tilde{g}'(y_j^*) < -\tilde{g}'(0)$. 
According to Theorem \ref{theorem:ne_condition1}, we have
$-w_jf(x_j^*)\tilde{g}'(y_j^*) = -w_if(x_i^*)\tilde{g}'(y_i^*)$. Then, we
can conclude $x_j^*>x_i^*$ such that $f'(x_j^*) < f'(x_i^*)$. 
The KKT condition in 
Equation~\eqref{eq:theorem2_no2} shows 
$w_jf'(x_j^*)\tilde{g}(y_j^*) > w_if'(x_i^*)\tilde{g}(0)$. 
Because $f'(x_j^*) < f'(x_i^*)$, $w_i > w_j$ and $\tilde{g}(y_j^*) < \tilde{g}(0)$,
the inequality does not hold. 
Hence, $\SD$  attacks $K_\SD$ with the descending order of the weights.

For the claim $K_\SA \geq K_\SD$, this can be inferred from our
preceding analysis since $\SD$
will not allocate resources to an target without being attacked by $\SA$ when $f(0) = 0$. \done

\subsection*{Proof of Lemma \ref{lemma:monotonicity}}

\noindent \textbf{Proof:} To search $(\lambda, \rho)$, we need to consider 
three different cases step by step: 1) $\lambda{>}0$ and $\rho{=}0$, 
2) $\lambda{=}0$ and $\lambda{>}0$ and 3) $\lambda{>}0$ and $\rho{>}0$.
Here, the change of $\lambda$ and $\rho$ does not alter $K_\SA$ and $K_\SD$ at the NE.
Recall that $h_\SA(\cdot)$ is a decreasing function and $h_\SD(\cdot)$ is an increasing function.
For simplicity, we let $\uparrow$ denote ``increase'' and let $\downarrow$ denote ``decrease''. 
The symbol $\Rightarrow$ denotes ``give rise to''.

\emph{Step 1: $\lambda{>}0$ and $\rho{=}0$}. 
When $\lambda$ $\uparrow$, $x_i^*$ $\downarrow$ for $K_\SD{<}i\leq K_\SA$.
For $1{\leq}i{\leq}K_\SD$, there are two possibilities, $x_i^*\uparrow$ or $x_i^* \downarrow$. 
In what follows, we will show that $x_i^*$ is strictly decreasing. 

We assume that $x_i^*$ $\uparrow$ as $\lambda \uparrow$. According to Eqs. \eqref{eq:xy_ne1} and 
\eqref{eq:xy_ne2}, we have the following relationships for all $1{\leq}i{\leq}K_\SD$:
$$\lambda \uparrow \Rightarrow x_i^* \uparrow \Rightarrow f(x_i^*) \uparrow \Rightarrow 
\frac{-\hat{c}}{w_if(x_i^*)} \uparrow \Rightarrow h_\SD(\frac{-\hat{c}}{w_if(x_i^*)}) \uparrow $$
$$\Rightarrow y_i^* \uparrow \Rightarrow \tilde{g}(y_i^*) \downarrow \Rightarrow \frac{c{+}\lambda}{w_i\tilde{g}(y_i^*)} \uparrow \Rightarrow h_\SA(\frac{c{+}\lambda}{w_i\tilde{g}(y_i^*)}) \downarrow \Rightarrow x_i^* \downarrow$$
which causes a self contradiction. Therefore, as $\lambda$ increases, $x_i^*$ cannot increases. 
It is easy to validate that $x_i^*$ cannot remain the same. Thus, $x_i^*$ is a strictly decreasing 
function of $\lambda$. According to the slackness condition, 
there has $\sum_{i=1}^{K_\SA}x_i^* = X_\SA$. If there exists a feasible $\lambda$ to satisfy this 
equality, $\lambda$ should be unique. A bisection algorithm can find the solution.

\emph{Step 2: $\lambda{=}0$ and $\rho{>}0$}. We assume that $y_i^* \uparrow$ 
when $\rho \uparrow$. Then, the following relationship holds:
$$\rho \uparrow \Rightarrow y_i^* \uparrow \Rightarrow \tilde{g}(y_i^*) \downarrow \Rightarrow \frac{c}{w_i\tilde{g}(y_i^*)} \uparrow \Rightarrow h_\SA(\frac{c}{w_i\tilde{g}(y_i^*)}) \downarrow $$
$$\Rightarrow x_i^* \downarrow \Rightarrow f(x_i^*) \downarrow \Rightarrow \frac{-(\hat{c}{+}\rho)}{w_if(x_i^*)} \downarrow \Rightarrow h_\SD(\frac{-(\hat{c}{+}\rho)}{w_if(x_i^*)}) \downarrow \Rightarrow y_i^* \downarrow$$
which contradicts to the assumption. Similarly, we can show that $y_i^*$ cannot remain unchanged.
 Therefore, when $\rho$ increases, $y_i^*$ is strictly decreasing for all $1{\leq}i{\leq}K_\SD$.
 The slackness condition gives rise to $\sum_{i=1}^{K_\SD}y_i^* = Y_\SD$.
 Then, we can use the bisection algorithm to find $\rho$ if it exists.

\emph{Step 3: $\lambda{>}0$ and $\rho{>}0$}. We consider two cases: $K_\SA=K_\SD$ and 
$K_\SA >K_\SD$. 

Recall that the implicit function $\rho_1(\lambda)$ is obtained from Eqs. \eqref{eq:slack1}\eqref{eq:xy_ne1}\eqref{eq:xy_ne2} and the implicit function $\rho_2(\lambda)$ is obtained from Eqs. \eqref{eq:slack2}\eqref{eq:xy_ne1}\eqref{eq:xy_ne2}.

\emph{Step 3.1 $K_\SA{=}K_\SD$}. When $\lambda$ increases, there are two cases due
to the constraint $\sum_{i{=}1}^{K_\SD} x_i^*{=}X_\SA$. 
One is that $x_i^*$ does not change for all $1{\leq}i{\leq}K_\SD$. The other is that 
there exist two targets $B_i$ and $B_j$  ($1{\leq}i,j{\leq}K_\SD$) in which $x_i^*$ increases and 
$x_j^*$ decreases. 

If $x_i^*$ does not change for all $1{\leq}i{\leq}K_\SD$, the following relationships hold
$$\lambda \uparrow \Rightarrow \tilde{g}(y_i^*) \uparrow \Rightarrow y_i^* \downarrow, 
\;\; \forall 1{\leq} i {\leq} K_\SD.$$
Because of $\rho{>}0$, there must have $\sum_{i{=}1}^{K_\SD} y_i^*{=}Y_\SD$, which contradicts
to the conclusion $y_i^*$ decreases for all $1{\leq} i {\leq} K_\SD$. Therefore, the case that 
$x_i^*$ ($1{\leq}i{\leq}K_\SD$) does not change is not true. 
 
We next turn to the second case that $x_i^*$ increases and $x_j^*$ decreases 
when $\lambda$ increases. The following relationships hold
$$\lambda \uparrow \Rightarrow x_i^* \uparrow \Rightarrow f(x_i^*) \uparrow \Rightarrow -\frac{1}{w_if(x_i^*)} \uparrow .$$
If $\rho$ increases or remains the same, we continue the induction by
$$\frac{-1}{w_if(x_i^*)} \uparrow \Rightarrow \frac{-(c{+}\rho)}{w_if(x_i^*)} \uparrow \Rightarrow
 h_\SD(\frac{-(c{+}\rho)}{w_if(x_i^*)}) \uparrow \Rightarrow y_i^* \uparrow $$
 $$\Rightarrow \tilde{g}(y_i^*) \downarrow \Rightarrow \frac{c{+}\lambda}{w_i\tilde{g}(y_i^*)} 
 \uparrow \Rightarrow h_\SA(\frac{c{+}\lambda}{w_i\tilde{g}(y_i^*)}) \downarrow \Rightarrow x_i^* \downarrow.$$
The condition $x_i^* \uparrow$ contradicts to the conclusion $x_i^*\downarrow$. Therefore, 
$\rho$ must decreases when $\lambda$ increases. In a word, $\rho_1(\lambda)$ is a 
strictly decreasing function.

According to the slackness condition in Eq. \eqref{eq:slack2}, there has 
$\sum_{i{=}1}^{K_\SD}y_i^* {=} Y_\SD$. When $\lambda$ increases, there are also two cases 
w.r.t. $y_i^*$. One is that $y_i^*$ does not change for $1{\leq} i {\leq} K_\SD$.
The other is that there exist two targets $B_i$ and $B_j$ ($1{\leq} i,j{\leq} K_\SD$) in which 
$y_i^*$ increases and $y_j^*$ decreases. 

If $y_i^*$ does not change for $1{\leq} i {\leq} K_\SD$, the following relationships hold
$$\lambda \uparrow \Rightarrow \frac{c{+}\lambda}{w_i\tilde{g}(y_i^*)} \uparrow \Rightarrow
h_\SA(\frac{c{+}\lambda}{w_i\tilde{g}(y_i^*)}) \downarrow \Rightarrow x_i^* \downarrow \Rightarrow
 \frac{-1}{w_if(x_i^*)} \downarrow.$$
 Because $y_i^*$ does not change, $\rho$ must increase. 
 
 For the second case, when $y_i^*$ increases, we obtain the following relationships
 $$\lambda_i^* \uparrow \Rightarrow \tilde{g}(y_i^*)\downarrow \Rightarrow \frac{c{+}\lambda}{w_i\tilde{g}(y_i^*)}
 \uparrow \Rightarrow h_\SD(\frac{c{+}\lambda}{w_i\tilde{g}(y_i^*)}) \downarrow $$
 $$\Rightarrow x_i^* \downarrow \Rightarrow f(x_i^*) \uparrow \Rightarrow \frac{-1}{w_if(x_i^*)}
 \downarrow.$$
 If $\rho$ decreases or remains the same, there must have
 $$\frac{-1}{w_if(x_i^*)} \downarrow \Rightarrow \frac{-(\hat{c}{+}\rho)}{w_if(x_i^*)}\downarrow
 \Rightarrow h_\SD(\frac{-(\hat{c}{+}\rho)}{w_if(x_i^*)})\downarrow \Rightarrow y_i^*\downarrow,$$
 which contradicts to the condition $y_i^*\uparrow$. Hence, $\rho$ must increase in this case. 
As a consequence, the implicit function $\rho_2(\lambda)$ is a strictly increasing function.
 
\emph{Step 3.2 $K_\SA{>}K_\SD$}. The slackness condition in Eq. \eqref{eq:slack1} is expressed as
\begin{eqnarray}
X_\SA = \sum_{i{=}1}^{K_\SD}h_\SA(\frac{c{+}\lambda}{w_i\tilde{g}(y_i^*(\lambda,\rho))})  + \sum_{i{=}K_\SD{+}1}^{K_\SA}h_\SA(\frac{c{+}\lambda}{w_i\tilde{g}(0)}).
\end{eqnarray}
When $\lambda$ increases, the expression $h_\SA(\frac{c{+}\lambda}{w_i\tilde{g}(0)})$ is strictly 
decreasing for $K_{\SD{+}1}{\leq}i{\leq}K_\SA$. This implies that $x_i^*$ decreases for 
$K_{\SD{+}1}{\leq}i{\leq}K_\SA$. Due to the constraint $\sum_{i{=}1}^{K_\SA}x_i^*{=} X_\SA$, 
$x_i^*$ increases in at least one target $B_i$ for $1{\leq}i{\leq}K_\SD$. 
In other word, the case that $x_i^*$ does not change with the increase of $\lambda$ does not happen. 
Then, following the analysis in the \emph{Step 3.1}, we can see that 
$\rho_1(\lambda)$ is a strictly decreasing function and $\rho_2(\lambda)$ is a strictly increasing 
function.

This concludes the proof. \done

\subsection*{Proof of Lemma \ref{lemma:neproperty}}

\noindent \textbf{Proof:} 
We prove this lemma by contradiction. 
When the both players allocate resource to targets
$B_i$ and $B_j$ at the NE, there exists
\begin{eqnarray}
\frac{w_i}{w_j}\frac{f(x_i^*)}{f(x_j^*)}\frac{\tilde{g}'(y_i^*)}{\tilde{g}'(y_j^*)} = \frac{w_i}{w_j}\frac{f'(x_i^*)}{f'(x_j^*)}\frac{\tilde{g}(y_i^*)}{\tilde{g}(y_j^*)} = 1.
\label{eq:property_proof_no1}
\end{eqnarray}
If $y_i^*<y_j^*$, the following inequality holds
\begin{eqnarray}
\tilde{g}'(y_i^*) < \tilde{g}'(y_j^*) < 0\nonumber
\label{eq:property_proof_no2}
\end{eqnarray}
because $g(\cdot)$ is strictly convex. The above inequality yields $\frac{\tilde{g}'(y_i^*)}{\tilde{g}'(y_j^*)}>1$. 
Combined with Eq.\eqref{eq:property_proof_no1}, we 
obtain $\frac{f(x_i^*)}{f(x_j^*)} < 1$. Since 
$f(\cdot)$ is strictly increasing and strictly concave, 
there have $x_i^*<x_j^*$ and $0<f'(x_j^*) <f'(x_i^*)$.
Then, we can conclude 
\begin{eqnarray}
\frac{w_i}{w_j}\frac{f'(x_i^*)}{f'(x_j^*)}\frac{\tilde{g}(y_i^*)}{\tilde{g}(y_j^*)} > 1, \noindent
\end{eqnarray}
which contradicts to Eq.\eqref{eq:property_proof_no1}.
Therefore, if $\SD$ allocates resource to targets $B_i$
and $B_j$, ($i<j$), at the NE, there must have $y_i^* > y_j^*$.

Eq.\eqref{eq:property_proof_no1} can be rewritten as
\begin{eqnarray}
\frac{f(x_i^*)}{f'(x_i^*)}\frac{\tilde{g}'(y_i^*)}{\tilde{g}(y_i^*)} = \frac{f(x_j^*)}{f'(x_j^*)}\frac{\tilde{g}'(y_i^*)}{\tilde{g}(y_j^*)}. 
\label{eq:property_proof_no5}
\end{eqnarray}
When $\frac{\tilde{g}'(y)}{\tilde{g}(y)}$ is a constant, there exists
$\frac{f(x_i^*)}{f'(x_i^*)} = \frac{f(x_j^*)}{f'(x_j^*)}$. If $x_i^*>x_j^*$, there have $f(x_i^*)> f(x_j^*)$ and $0\leq f'(x_i^*)<f(x_j^*)$. This gives 
rise to the inequality 
$\frac{f(x_i^*)}{f'(x_i^*)} > \frac{f(x_j^*)}{f'(x_j^*)}$, which contradicts to the above equality.
It is also easy to show that the relationship 
$x_i^*<x_j^*$ also contradicts to the above equality. 
Hence, we obtain $x_i^*=x_j^*$. We next suppose that  $\frac{\tilde{g}'(y)}{\tilde{g}(y)}$ is an increasing function of $y$.
Given $y_i^*>y_j^*$ for $1\leq i{<}j{\leq} K_\SD$, we 
obtain $\frac{\tilde{g}'(x_j^*)}{\tilde{g}(x_j^*)} < \frac{\tilde{g}'(x_i^*)}{\tilde{g}(x_i^*)} < 0$. Then, eq.\eqref{eq:property_proof_no5}
yields $\frac{f(x_i^*)}{f'(x_i^*)} > \frac{f(x_j^*)}{f'(x_j^*)}$, or equivalently $x_i^*>x_j^*$. Similarly,
when $\frac{\tilde{\tilde{g}}'(y)}{\tilde{g}(y)}$ is a strictly decreasing 
function of $y$, there must have $x_i^*<x_j^*$. \done

\subsection*{Proof of Lemma \ref{lemma:linear_neproperty1}}

\noindent \textbf{Proof:} This lemma is proved by contradiction. We consider even more
general functions: $f(x)=b_1x$ and $g(y)=b_2-b_3y$. In the intrusion detection game, 
we let $b_1=b_2=1$ and $b_3>0$. 

i). We assume $x_i^* {=} 0$ and $x_j^* {>} 0$ at the NE for two targets $B_i$ and $B_j$ 
with $w_i{>}w_j$. The best response of $\SD$ must satisfy $y_i^*{=}0$.
Then, the following inequality holds
\begin{eqnarray}
\frac{dU_\SA}{dx_i}|_{x_i{=}x_i^*} {=} w_ib_1b_2{-}c {>}w_jb_1b_2{-}w_jb_2b_3y_j^*{-}c {=} \frac{dU_\SA}{dx_j}|_{x_j{=}x_j^*}. \nonumber
\end{eqnarray}
$\SA$ obtains a higher profit if he transfers the resource on $B_j$ to $B_i$. Thus, it is not a NE. 

We further assume $y_i^*{=}0$ and $y_j{>}0$ at the NE for two targets $B_i$ and $B_j$ 
with $w_i{>}w_j$. The marginal profits on $B_i$ and $B_j$ satisfy 
\begin{eqnarray}
w_ib_1b_3x_i^*< w_jb_1b_3 x_j^*. \nonumber
\end{eqnarray}
The above inequality gives rise to $x_i^*{<}x_j^*$ because of $w_i{>}w_j$.
When $y_i^*{=}0$ and $y_j^*{>}0$, the marginal profits of $\SA$ on $B_i$ and $B_j$ satisfy 
$\frac{dU_\SA}{dx_i}|_{x_i{=}x_i^*}  > \frac{dU_\SA}{dx_i}|_{x_j{=}x_j^*}$. Then, $\SA$
has a larger utility if he moves the resource on $B_j$ to $B_i$. This contradicts to the claim 
$x_i^*{<}x_j^*$. Thus, it is not a NE. 

To sum up, $\SA$ allocates resources to the top $K_\SA$ targets and $\SD$ allocates resources to 
the top $K_\SD$ targets. It is also very intuitive to validate $K_\SD{<}K_\SA$.

ii). We assume $K_\SA {>} K_\SD {+}1$ at the NE. Let $B_i$ and $B_j$ be two targets for 
$K_\SD{<}i,j{<}N$. The marginal profits of $\SA$ on $B_i$ and $B_j$ satisfy
\begin{eqnarray}
\frac{dU_\SA}{dx_i} = w_ib_1b_2{-}c, \;\; \frac{dU_\SA}{dx_j} = w_jb_1b_2 -c. \nonumber
\end{eqnarray}
Because of $w_i{\neq} w_j$, $\SA$ can obtain a larger utility by aggregating the resources to the more
profitable target. Thus, it is not a NE. To sum up, $K_\SA$ and $K_\SD$ must satisfy
\begin{eqnarray}
K_\SD \leq K_\SA \leq K_\SD {+} 1. \nonumber
\end{eqnarray}
This concludes the proof.

\subsection*{Proof of Theorem \ref{theorem:ne_linear}}

\noindent \textbf{Proof:} 
The proof utilizes the conclusions of
lemma \ref{lemma:linear_neproperty1}. According to the properties of the NE, there have
\begin{eqnarray}
&&\!\!\!\!\!\!\!\!\!\!\!\!\!\!\!\!\!\!\!\!w_{i}b_1b_2 {-}w_ib_1b_3y_i^* {-}c = \lambda \geq 0, \quad \forall \;\; 1{\leq}i{\leq} K_\SA;\label{eq:proof_theorem3_no1} \\
&&\!\!\!\!\!\!\!\!\!\!\!\!\!\!\!\!\!\!\!\! w_{K_\SD}b_1b_2 -  w_{K_\SD}b_1b_3 y_{K_\SD}^*\geq  w_{K_\SD{+}1}b_1b_2 \;\; \textrm{if} \;\; K_\SD{<} N; \label{eq:proof_theorem3_no2}\\
&&\!\!\!\!\!\!\!\!\!\!\!\!\!\!\!\!\!\!\!\!w_i b_1b_3x_i^* - \hat{c} = \rho \geq 0, \quad \forall \;\; 1{\leq}i{\leq} K_\SD;\label{eq:proof_theorem3_no3} \\
&&\!\!\!\!\!\!\!\!\!\!\!\!\!\!\!\!\!\!\!\!w_i b_1b_3x_{K_\SD{+}1}^* - \hat{c} \leq\rho , \;\; \textrm{if} \;\; K_\SD{<} N.\label{eq:proof_theorem3_no4} 
\end{eqnarray}
Here, Eq.\eqref{eq:proof_theorem3_no1} means that the marginal utilities of 
$\SA$ are non-negative and 
are the same on the top $K_\SA$ targets. Eq. \eqref{eq:proof_theorem3_no2} means that the 
marginal utility of $\SA$ on any top $K_\SD$ target is larger than that on target $B_{K_\SD{+}1}$.
This guarantees the condition $K_\SD {\leq} K_\SA {\leq} K_\SD{+}1$. 
Eq. \eqref{eq:proof_theorem3_no3} ensures that $\SD$ allocates positive resources to the top 
$K_\SD$ targets. Eq. \eqref{eq:proof_theorem3_no4} means that $\SD$ does not allocate 
resources to $B_{K_\SD{+}1}$. The above conditions give rise to the solution to the NE,
\begin{eqnarray}
x_i^* \!\!\!\!\!&&\!\!\!\!\!  \left\{\begin{matrix}
=\frac{\hat{c}+\rho}{w_ib_1b_3}, & \forall \; i\leq K_\SD \\
\;\leq\frac{\hat{c} + \rho}{w_{K_\SD{+}1}b_1b_3}, & \quad i {=} K_\SD{+}1 \; (K_\SD{<}N)
\end{matrix}\right., \\
y_i^* \!\!\!\!\!&&\!\!\!\!\!  \left\{\begin{matrix}
=\frac{b_2}{b_3} {-} \frac{c{+}\lambda}{w_ib_1b_3}, & \forall \; i\leq K_\SD\\
\;\leq (1-\frac{w_{K_\SD{+}1}}{w_{i}})\frac{b_2}{b_3}, &\quad \forall \; i{\leq}K_\SD \;(K_\SD{<}N)
\end{matrix}\right..
\end{eqnarray}

Before commencing the analysis, we recall the following
notations: $P_\SA(k)$ and $P_\SD(k)$ are defined as 
$P_\SA(0) {:=}0$, 
$P_\SA(k){:=}\sum\nolimits_{i{=}1}^k\frac{\hat{c}}{w_ib_1b_3}, \; \forall \; 1{\leq}k{\leq}N$; $P_\SD(1) {=} 0$,  $P_\SD(k){:=}\sum\nolimits_{i{=}1}^{k{-}1}\frac{b_2}{b_3}(1{-}\frac{w_{k}}{w_i})$, and $P_\SD(N{+}1)  {:=} \frac{b_2}{b_3}N{-}\sum\nolimits_{i{=}1}^{N}\frac{c}{w_ib_1b_3}$.

\noindent i). We first prove the following claim via three steps:
\begin{itemize}
\item $P_\SA(k) \!{<}X_\SA \!{<}\! P_\SA(k{+}1)$ and $Y_\SD \!{>} P_\SD(k{+}1)$ for $0{\leq}\!k{\leq}\!N{-}1$.
The NE is uniquely determined by
{
\setlength{\abovedisplayskip}{5pt}
\setlength{\belowdisplayskip}{5pt}
\begin{eqnarray}
\!\!\!\!\!x_i^*\!\!\!&=&\!\!\!\left\{\begin{matrix}
\;\frac{\hat{c}}{w_ib_1b_3}, & \forall \; i\leq k \\
\;X_\SA{-}\sum_{j=1}^k \frac{\hat{c}}{w_jb_1b_2}, & i {=} k{+}1 \\
\;0, & \forall \; i {>} k{+}1 
\end{matrix}\right. 
\label{eq:proof_linearne1}\\
\!\!\!\!\!y_i^* \!\!\!&=&\!\!\! \left\{\begin{matrix}
\; (1-\frac{w_{k{+}1}}{w_i})\frac{b_2}{b_3} , & \forall \; i{\leq} k \\
\;0, & \forall \; i > k 
\end{matrix}\right. .
\label{eq:proof_linearne2}
\end{eqnarray}
}
\end{itemize}

\noindent \emph{Step 1.1} $K_\SD$ cannot be less than $k$

We assume $K_\SD{<}k$. 
If $\lambda {>} 0$, $\SA$ allocates all of his resources on the 
top $K_\SA$ targets, that is,
\begin{eqnarray}
X_\SA = \sum\nolimits_{i{=}1}^{K_\SD}\frac{\hat{c}{+}\rho}{w_ib_1b_3}  + x_{K_\SD{+}1}^*
 {\leq}  \sum\nolimits_{i{=}1}^{K_\SD{+}1}\frac{\hat{c}{+}\rho}{w_ib_1b_3}.
\label{eq:proof_theorem3_no10}
\end{eqnarray}
Because of $\sum\nolimits_{i{=}1}^{k}\frac{\hat{c}}{w_ib_1b_3} {<}X_\SA  
{<}  \sum\nolimits_{i{=}1}^{k{+}1}\frac{\hat{c}}{w_ib_1b_3}$, there has 
\begin{eqnarray}
\sum\nolimits_{i{=}1}^{k}\frac{\hat{c}}{w_ib_1b_3} {<} \sum\nolimits_{i{=}1}^{K_\SD{+}1}\frac{\hat{c}{+}\rho}{w_ib_1b_3}.
\label{eq:proof_theorem3_no11}
\end{eqnarray}
Due to the condition $K_\SD{<}k$, the above inequality gives rise to $\rho{>}0$, which means that the marginal utility of $\SD$ is positive. Thus, $\SD$ allocates all the resources
to the top $K_\SD$ targets. According to the expression of NE, the total resources allocated by 
$\SD$ on $K_\SD$ targets satisfy
\begin{eqnarray}
Y_\SD \leq \sum\nolimits_{i{=}1}^{K_\SD}(1-\frac{w_{K_\SD{+}1}}{w_i})\frac{b_2}{b_3}.
\label{eq:proof_theorem3_no12}
\end{eqnarray}
This contradicts to the condition $Y_\SD {>} \sum_{j=1}^{k}(1{-}\frac{w_{k{+}1}}{w_j})\frac{b_2}{b_3}$
when $K_\SD{<} k$.

If $\lambda{=}0$, the marginal utility on any target $B_i$ that has no resource of $\SD$ 
is given by $w_ib_1b_2 - c>0$ ($i{>}K_\SD$). 
$\SA$ obtains a larger utility by shifting resources to 
any unprotected target, which is a feasible NE. 
Therefore, $K_\SD$ cannot be less than $k$.

\noindent \emph{Step 1.2} $K_\SD$ cannot be larger than $k$

We assume $K_\SD > k$. The total amount of resources used by $\SA$ at the NE is given by
\begin{eqnarray}
\sum\nolimits_{i{=}1}^{K_\SD{+}1}x_i^* {=} \sum\nolimits_{i{=}1}^{K_\SD} \frac{\hat{c}{+}\rho}{w_ib_1b_3}
{+} x_{K_\SD{+}1}^* {\geq} \sum\nolimits_{i{=}1}^{K_\SD} \frac{\hat{c}}{w_ib_1b_3}.\nonumber
\label{eq:proof_theorem3_no13}
\end{eqnarray}
Due to the conditions $K_\SD {>} k$ and $X_\SA {<} \sum_{i=1}^{k{+}1}\frac{\hat{c}}{w_ib_1b_3}$, 
we obtain $\sum\nolimits_{i{=}1}^{K_\SD{+}1}x_i^* {>} X_\SA$, which is not true.
Hence, $K_\SD$ cannot be larger than $k$.

\noindent \emph{Step 1.3} $K_\SD$ is equal to $k$

In the above analysis, we observe that $\lambda$ must satisfy
\begin{eqnarray}
\lambda \geq w_{k{+1}} b_1b_2 - c,
\label{eq:proof_theorem3_no14}
\end{eqnarray}
given the condition $k{<}N$. Otherwise, $\SA$ can perform better by moving 
the resources to the $(k{+}1)^{th}$ target.  Since $\lambda{>} 0$, $\SA$ fully utilizes his resources.

We then consider the value of $\rho$. When $\rho{>}0$, $\SD$ allocates all the resources to the 
top $k$ targets. This yields
\begin{eqnarray}
Y_\SD =\frac{b_2}{b_3}k - \sum\nolimits_{i{=}1}^{k}\frac{c{+}\lambda}{w_ib_1b_3} .\label{eq:proof_theorem3_no15}
\end{eqnarray}
Submitting \eqref{eq:proof_theorem3_no14} to \eqref{eq:proof_theorem3_no15}, we obtain 
the condition 
$Y_\SD{\leq}\sum\nolimits_{i{=}1}^{k}(1{-}\frac{w_{k{+}1}}{w_i})\frac{b_2}{b_3}$. This contradicts 
to the initial condition $Y_\SD{>} \sum\nolimits_{i{=}1}^{k}(1{-}\frac{w_{k{+}1}}{w_i})\frac{b_2}{b_3}$.
Hence, $\rho$ cannot be greater than 0. 
When $\rho=0$, the NE strategies of $\SA$ and $\SD$ can be easily solved by \eqref{eq:proof_linearne1} and \eqref{eq:proof_linearne2}.

ii) We next prove the second claim. 

\begin{itemize}
\item $P_\SD(k) {<} Y_\SD {<} P_\SD(k{+}1)$ and $X_\SA {>} P_\SA(k)$ for $1{\leq}k{\leq}N$.
The NE is uniquely determined by
{
\setlength{\abovedisplayskip}{5pt}
\setlength{\belowdisplayskip}{5pt}
\begin{eqnarray}
\!\!\!\!\!x_i^*\!\!&=&\!\!\left\{\begin{matrix}
\;(\sum_{j=1}^{k}\frac{w_i}{w_j})^{-1}X_\SA, & \forall \; i\leq k \\
\;0, & \forall \; i > k
\end{matrix}\right. 
\label{eq:proof_linearne3}\\
\!\!\!\!\!y_i^* \!\!&=&\!\! \left\{\begin{matrix}
 (\sum\nolimits_{j=1}^{k}\frac{w_i}{w_j})^{-1}
\big(Y_\SD  {-} \frac{b_2}{b_3}k \big) {+} \frac{b_2}{b_3} , & \!\!\forall  i{\leq} k \\
0, & \!\!\forall  i {>} k 
\end{matrix}\right. .
\label{eq:proof_linearne4}
\end{eqnarray}
}
\end{itemize}

\noindent \emph{Step 2.1 $K_\SD$ cannot be less than $k$}

We assume $K_\SD{<}k$. If $\lambda{>}0$, we obtain the condition $\rho{>}0$ 
following the expression in \eqref{eq:proof_theorem3_no10}. This means that 
$\SD$ allocates $Y_\SD$ resources to the top $K_\SD$ targets. Then, there has the following 
inequality at the NE
\begin{eqnarray}
Y_\SD \leq \sum\nolimits_{i{=}1}^{K_\SD}(1-\frac{w_{K_\SD{+}1}}{w_i})\frac{b_2}{b_3}.
\label{eq:proof_theorem3_no20}
\end{eqnarray}
Note that the feasible region of $Y_\SD$ is $Y_\SD>\sum_{j=1}^{k{-}1}(1{-}\frac{w_{k}}{w_j})\frac{b_2}{b_3}$. Because of $K_\SD{<}k$, there has
\begin{eqnarray}
Y_\SD {>} \!\!\!\sum_{j=1}^{k{-}1}(1{-}\frac{w_{k}}{w_j})\!\frac{b_2}{b_3} {\geq}\!\!
\sum_{j=1}^{K_\SD}(1{-}\frac{w_{k}}{w_j})\!\frac{b_2}{b_3} {\geq} \!\!\sum_{j=1}^{K_\SD}(1{-}\frac{w_{K_\SD{+}1}}{w_j})\!\frac{b_2}{b_3}.
\label{eq:proof_theorem3_no21}
\end{eqnarray}
The inequality \eqref{eq:proof_theorem3_no20} contradicts to \eqref{eq:proof_theorem3_no21}, 
which means that $\lambda$ cannot be greater than 0. 

If $\lambda{=}0$, all the resources of $\SA$ will be moved to target $B_{K_\SD{+}1}$. Then,
this is not a NE. Therefore, $K_\SD$ cannot be less than $k$.

\noindent \emph{Step 2.2 $K_\SD$ cannot be larger than $k$}

We assume $K_\SD{>}k$ with conditioned on $k{<}N$. The total amount of resources used by $\SD$ at the NE satisfy
\begin{eqnarray}
\sum\nolimits_{i=1}^{K_\SD}y_i^* {=} \sum\nolimits_{i=1}^{K_\SD}(\frac{b_2}{b_3}{-}\frac{c{+}\lambda}{w_ib_1b_3})
\label{eq:proof_theorem3_no22}
\end{eqnarray}
There must have $\lambda{\geq} w_{K_\SD{+}1}b_1b_2{-}c$ 
if $\SA$ allocates positive resources to
target $B_{K_\SD}$. Considering the additional condition $K_\SD{>}k$, 
Eq. \eqref{eq:proof_theorem3_no22} yields
\begin{eqnarray}
\sum\nolimits_{i=1}^{K_\SD}y_i^* {\geq} \sum\nolimits_{i=1}^{K_\SD}(1{-}\frac{w_{K_\SD{+}1}}{w_i})\frac{b_2}{b_3} {>} \sum\nolimits_{i=1}^{k}(1{-}\frac{w_{k{+}1}}{w_i}).
\label{eq:proof_theorem3_no23}
\end{eqnarray}
The resource limit of $\SD$ should satisfy $Y_\SD{\geq}\sum\nolimits_{i=1}^{K_\SD}y_i^*$.
However, the inequality \eqref{eq:proof_theorem3_no23} contradicts to the condition 
$Y_\SD < \sum_{j=1}^{k}(1{-}\frac{w_{k{+}1}}{w_j})\frac{b_2}{b_3}$. Therefore, $K_\SD$ 
cannot be larger than $k$.

\noindent \emph{Step 2.3 $K_\SD$ is equal to $k$}

We consider two scenarios separately, $k{<}N$ and $k=N$. 

If $k{<}N$, there must have $\lambda\geq w_{k{+}1}b_1b_2-c$ according to Eq. \eqref{eq:proof_theorem3_no13}. 
If the equality $\lambda= w_{k{+}1}b_1b_2-c$ holds, the total amount of resources used by $\SD$
at the NE is given by $\sum\nolimits_{i{=}1}^{k}(1{-}\frac{w_{k{+}1}}{w_i})\frac{b_2}{b_3}$. This 
contradicts to the range of $Y_\SD$. Hence, there only has $\lambda{>} w_{k{+}1}b_1b_2{-}c$, 
which means that both $\SA$ and $\SD$ allocate positive resources to $k$ targets. 
Since $\lambda{>}w_{k{+}1}b_1b_2{-}c$, there exists $X_\SA{=}\sum_{i{=}1}^{k}x_i^* = \sum_{i{=}1}^{k}\frac{\hat{c}{+}\rho}{w_ib_1b_3}$. Because $X_\SA{>}\sum_{i{=}1}^{k}\frac{\hat{c}}{w_ib_1b_3}$, 
$\rho$ must be positive. Hence, by letting $X_\SA{=}\sum_{i{=}1}^{k}x_i^* $ and $Y_\SD{=}\sum_{i{=}1}^{k}y_i^* $, we can directly solve the NE as
\begin{eqnarray}
x_i^*\!\!\!\!\!\!&=&\!\!\!\!\!\!
\;(\sum_{j=1}^{k}\frac{w_i}{w_j})^{-1}X_\SA, \quad \forall \; i\leq k \label{eq:proof_theorem3_no26}\\
y_i^* \!\!\!\!\!\!&=&\!\!\!\!\!\! 
 (\sum\nolimits_{j=1}^{k}\frac{w_i}{w_j})^{-1}
\big(Y_\SD  {-} \frac{b_2}{b_3}k \big) {+} \frac{b_2}{b_3} ,  \quad \!\!\forall  i{\leq} k \label{eq:proof_theorem3_no27}
\end{eqnarray}

If $k{=}N$, there has $\lambda{\geq} 0$. Here, when $\lambda{=}0$, the total amount of resources 
utilized by $\SD$ at the NE is given by $\sum\nolimits_{i{=}1}^N y_i^* {=} \frac{b_2}{b_3}N {-} \sum_{i{=}1}^{N}\frac{c}{w_ib_1b_3}$. Because $Y_\SD < \frac{b_2}{b_3}N {-} \sum_{i{=}1}^{N}\frac{c}{w_ib_1b_3}$, there has $\sum\nolimits_{i{=}1}^N y_i^* {>} Y_\SD$, which is not true. Hence, 
$\lambda$ is always greater than 0. It is easy to conclude $\rho {>} 0$ since 
$X_\SA{>}\sum\nolimits\frac{c_i}{w_ib_1b_3}$ are fully utilized at the NE. 
Now we are clear that both $X_\SA$ and $Y_\SD$ 
are disposed on all $N$ targets. The NE can be computed in the same 
way as that in Eqs. \eqref{eq:proof_theorem3_no26} and \eqref{eq:proof_theorem3_no27}.

iii) We then prove the third claim. 

\begin{itemize}

\item $X_\SA {>} P_\SA(N)$ and $Y_\SD {>} P_\SD(N{+}1)$, the NE is given by 
\begin{eqnarray}
x_i^* = \frac{\hat{c}}{w_ib_1b_3} \;\; \textrm{and} \;\; \frac{b_2}{b_3} - \frac{c}{w_ib_1b_3} . 
\label{eq:proof_theorem3_no30}
\end{eqnarray}

\end{itemize}

To prove this claim, we only  need to show that $\lambda$ and $\rho$ are both 0 at the NE. We still 
prove it by contradiction. If $\lambda{>}0$, all the resources of $\SA$ are allocated to these $N$ 
targets. Because $X_\SA$ is larger than $\sum\nolimits_{i=1}^{N}\frac{\hat{c}}{w_ib_1b_3}$, $\rho$
must be positive in the marginal utility functions. As a countermeasure, $\SD$ allocates all the 
resources to defend these targets. However, after $\SD$ allocates all of his resources, the marginal
utilities of $\SA$ become negative due to $Y_\SD {>}\frac{b_2}{b_3}N{-}
\sum\nolimits_{i=1}^{N}\frac{c}{w_ib_1b_3}$. The best strategy of $\SA$ is to give up all the targets. 
Hence, either $\lambda$ and $\rho$ cannot be 0 at the NE.  The only possible NE must satisfy 
$\lambda{=}\rho{=}0$, which leads to the expression of the NE in Eq. \eqref{eq:proof_theorem3_no30}.

iv) We continue to prove the fourth claim.

\begin{itemize}
\item  $X_\SA {=} P_\SA(k)$ and $Y_\SD {\geq} P_\SD(k)$ for $1{\leq}k{\leq}N$. 
Denote $\tilde{Y}_\SD$ as any real value in the range
$[P_\SD(k), \min\{Y_\SD, P_\SD(k{+}1)\}]$. There exist 
multiple NEs given by 
{
\setlength{\abovedisplayskip}{5pt}
\setlength{\belowdisplayskip}{5pt}
\begin{eqnarray}
\!\!\!\!\!\!x_i^*\!\!\!\!&=&\!\!\!\!\left\{\begin{matrix}
\;\frac{\hat{c}}{w_ib_1b_3}, & \forall \;\; i\leq k \\
\;0, & \forall \;\;  k{+}1{\leq} i{\leq}N
\end{matrix}\right.  \label{eq:proof_linearne7}
\\
\!\!\!\!\!\!y_i^* \!\!\!\!&=&\!\!\!\! \left\{\begin{matrix}
\; \frac{b_2}{b_3} {+} (\sum\nolimits_{j{=}1}^k \frac{w_i}{w_j})^{-1}(\tilde{Y}_\SD {-} k\frac{b_2}{b_3}), & \forall \; i\leq k \\
\;0, & \forall \;i > k 
\end{matrix}\right.  \label{eq:proof_linearne8}
\end{eqnarray}
}
\end{itemize}

When $x_i^*$ is taken as $\frac{\hat{c}}{w_ib_1b_3}$ for $i{\leq}k$, the marginal utilities of $\SD$ 
are always 0 on the targets from $B_1$ to $B_k$. This means that $\SD$ cannot obtain a better 
utility by unilaterally changing his strategy. 
In this scenario, $\SA$ does not change his allocation strategy 
as long as his marginal utilities on the targets from $B_1$ to $B_k$ are the same and 
are non-negative. Let $\tilde{Y}_\SD$ be the total amount of resources utilized by $\SD$ 
at the NE. There must have 
\begin{eqnarray}
\sum\nolimits_{i{=}1}^k y_i^* = \sum\nolimits_{i{=}1}^k(\frac{b_2}{b_3} - \frac{c{+}\lambda}{w_ib_1b_3})
=\tilde{Y}_\SD.
\end{eqnarray}
Therefore, the strategy of $\SD$ is obtained by
\begin{eqnarray}
y_i^* = \frac{b_2}{b_3} {+} (\sum\nolimits_{j{=}1}^k \frac{w_i}{w_j})^{-1}(\tilde{Y}_\SD {-} k\frac{b_2}{b_3}), \;\; \forall \; 1{\leq}i{\leq}k
\end{eqnarray}
and $y_i^*=0$ for $i{>}k$. 
Note that $\tilde{Y}_\SD$ cannot be larger than $P_\SD({k{+}1})$. Otherwise, the marginal utilities 
of $\SA$ on $B_1$ to $B_k$ become negative such that $\SA$ gives up these targets.

v.) We finally prove the fifth claim. 

\begin{itemize}
\item  $Y_\SD {=} P_\SD(k)$ and $P_\SA(k{-}1) {\leq} X_\SA {\leq} P_\SA(k)$ for $2{\leq}k{\leq}N$. We denote $\tilde{X}_\SA$ in the range
$[P_{\SA}(k{-}1), X_\SA]$.  There exist multiple NEs given by 
{
\setlength{\abovedisplayskip}{5pt}
\setlength{\belowdisplayskip}{5pt}
\begin{eqnarray}
x_i^*\!\!\!&=&\!\!\!\left\{\begin{matrix}
\;(\sum\nolimits_{j{=}1}^k \frac{w_i}{w_j})^{-1}\tilde{X}_\SA, & \forall \; i\leq k \\
\;0, & \forall \; i > k{+}1 
\end{matrix}\right.  \label{eq:proof_linearne9}
\\
y_i^* \!\!\!&=&\!\!\! \left\{\begin{matrix}
\; (1{-}\frac{w_{k{+}1}}{w_i})\frac{b_2}{b_3} , & \forall \; i\leq k \\
\;0, & \forall \; i > k 
\end{matrix}\right.  \label{eq:proof_linearne10}
\end{eqnarray}
}
\end{itemize}

When $y_i^*$ is taken as $\frac{b_2}{b_3}(1{-}\frac{w_{k{+}1}}{w_i})$, the marginal utilities of 
$\SA$ on targets from $B_1$ to $B_k$ are all 0. Then, $\SA$ cannot improve his utility by 
individually changing his strategy. 
At the NE, the marginal utilities of $\SD$ on targets from $B_1$ to $B_k$ 
should be non-negative and identical. 
Let $\tilde{X}_\SA$ be the amount of resources used by $\SA$ at the NE. There exist
$w_ix_i^* {=} w_jx_j^* {>}0$ for all $i,j\leq k$ and $\sum\nolimits_{i{=}1}^kx_i^* {=} \tilde{X}_\SA$.
Hence, the NE strategy of $\SA$ is given by $x_i^*=(\sum\nolimits_{j{=}1}^k \frac{w_i}{w_j})^{-1}\tilde{X}_\SA$ for $i\leq k$ and $x_i^* = 0$ for $i>k$. 

This concludes the proof. \done

\subsection*{Proof of Theorem \ref{theorem:ne_coupled}}

The proof follows that of Theorem \ref{theorem:ne_condition1}. 
Let $\lambda$ and $\rho$ be the Lagrange multipliers of $\SA$ and $\SD$ respectively. 
Let $L_\SD(\by,\rho)$ be the Lagrange function of the defender $\SD$ that has
\begin{eqnarray}
L_\SD(\by,\rho) = -\sum\nolimits_{i=1}^{N} \frac{w_if(x_i)}{f(x_i)+g(y_i)} - \hat{c}\sum\nolimits_{i=1}^{N}y_i \nonumber\\
+\rho(Y_\SD-\sum\nolimits_{i=1}^Ny_i).
\end{eqnarray}
We take the derivative of $L_\SD(\by,\rho)$ over $y_i$ and obtain
\begin{eqnarray}
\frac{dL_\SD(\by,\rho)}{dy_i} = \frac{w_i f(x_i)g'(y_i)}{(f(x_i)+g(y_i))^2} - (\hat{c}+\rho), \quad \forall i.
\end{eqnarray}
Here, $L_\SD(\by,\rho)$ is optimized in two ways. If the above derivative is 0, there exists a non-zero 
resource allocation strategy, i.e. $y_i^*>0$. If the above derivative is less than 0, then $y_i^*$ 
is 0. Similarly, we can find the conditions for the attacker to maximize his utility.
For the sake of redundancy, we omit the detailed proof. \done

\subsection*{Proof of Lemma \ref{lemma:coupled_targets}}

\noindent\textbf{Proof:} According to Theorem \ref{lemma:existence}, 
these exists a unique NE with the proportion-form
breaching model. We next show by contradiction that $x_i$ cannot be 0 on any target $B_i$ at the NE. 
Suppose $x_i^*=0$ on target $B_i$. Then, there has $f(x_i^*)=0$ such that $y_i^*$ is 0.
When target $B_i$ is not protected by $\SD$, the best response of $\SA$ is to allocate an arbitrarily 
small amount of resources to this target. Hence, $(0,0)$ is not an equilibrium strategy for $\SA$ and $\SD$. 
Therefore, $\SA$ and $\SD$ allocate positive resources to all the targets at the NE. \done

\subsection*{Proof of Lemma \ref{lemma:coupled_strategy}}

\noindent\textbf{Proof:} Consider two targets $B_i$ and $B_j$ with $w_i>w_j$. 
The following equations hold at the NE.
\begin{eqnarray}
\frac{w_if'(x_i)g(y_i)}{(f(x_i)+g(y_i))^2} = \frac{w_jf'(x_j)g(y_j)}{(f(x_j)+g(y_j))^2} = c+\lambda; 
\label{eq:proof_proportion_form1}\\
\frac{w_if(x_i)g'(y_i)}{(f(x_i)+g(y_i))^2} = \frac{w_jf(x_j)g'(y_j)}{(f(x_j)+g(y_j))^2} = \hat{c}+\rho.
\label{eq:proof_proportion_form2}
\end{eqnarray}
The above equations yield the following relationship
\begin{eqnarray}
\frac{f'(x_i)}{f'(x_j)} \frac{f(x_j)}{f(x_i)} = \frac{g'(y_i)}{g'(y_j)} \frac{g(y_j)}{g(y_i)}.
\label{eq:proof_proportion_form3}
\end{eqnarray}
We prove this lemma by contradiction. Let us assume that there has $y_i<y_j$. 
Because $g(\cdot)$ is a concave and strictly increasing function, 
we have $g(y_i)<g(y_j)$ and $g'(y_i)>g'(y_j)$. The right hand of 
Eq.\eqref{eq:proof_proportion_form3} is greater than 1. Then, there must have 
$x_i<x_j$ in the left hand of Eq.\eqref{eq:proof_proportion_form3}.

We define two functions, $f_1(x,y)$ and $f_2(x,y)$, where
\begin{eqnarray}
f_1(x,y) = \frac{f'(x)g(y)}{(f(x){+}g(y))^2} \textrm{ and } f_2(x,y) = \frac{f(x)g'(y)}{(f(x){+}g(y))^2} .
\label{eq:proof_proportion_form4}
\end{eqnarray} 
We take the derivatives of $f_1(x,y)$ and $f_2(x,y)$ over $x$ and $y$ respectively. 
\begin{eqnarray}
\frac{\partial f_1}{\partial x} \!\!\!&=&\!\!\! g(y)\frac{f''(x)(f(x){+}g(y)){-}2(f'(x))^2}{(f(x)+g(y))^2} < 0;
\label{eq:proof_proportion_form5}\\
\frac{\partial f_2}{\partial y} \!\!\!&=&\!\!\! f(x)\frac{g''(x)(f(x){+}g(y)){-}2(g'(y))^2}{(f(x)+g(y))^2} < 0;
\label{eq:proof_proportion_form6}\\
\frac{\partial f_1}{\partial y} \!\!\!&=&\!\!\! f'(x)g'(y)\frac{f(x)-g(y)}{(f(x)+g(y))^2} ;
\label{eq:proof_proportion_form7}\\
\frac{\partial f_2}{\partial x} \!\!\!&=&\!\!\! f'(x)g'(y)\frac{g(y)-f(x)}{(f(x)+g(y))^2} .
\label{eq:proof_proportion_form8}
\end{eqnarray} 
The signs of $\frac{\partial f_1}{\partial y}$ and $\frac{\partial f_2}{\partial x}$ depend on whether 
$f(x)$ is greater than $g(y)$ or not. Meanwhile, $f_1(x,y)$ is a decreasing function of $x$ 
and $f_2(x,y)$ is an increasing function of $y$. 

To prove this lemma, we consider two cases, $f(x_i) > g(y_i)$ and $f(x_i)<g(y_i)$. 

\noindent \textit{Case 1: $f(x_i) {>} g(y_i)$.} Because there has $f(x_j){>}f(x_i){>}g(y_i)$, we obtain 
\begin{eqnarray}
\frac{f(x_i)g'(y_i)}{(f(x_i){+}g(y_i))^2}  > \frac{f(x_j)g'(y_i)}{(f(x_j){+}g(y_i))^2}  .
\label{eq:proof_proportion_form9}
\end{eqnarray}
Since $f_2(x,y)$ is strictly decreasing w.r.t. $y$, there yields
\begin{eqnarray}
\frac{f(x_j)g'(y_i)}{(f(x_j){+}g(y_i))^2}  > \frac{f(x_j)g'(y_j)}{(f(x_j){+}g(y_j))^2} .
\label{eq:proof_proportion_form10}
\end{eqnarray}
Submitting \eqref{eq:proof_proportion_form10} to \eqref{eq:proof_proportion_form9}, we have 
\begin{eqnarray}
\frac{f(x_i)g'(y_i)}{(f(x_i){+}g(y_i))^2} > \frac{f(x_j)g'(y_j)}{(f(x_j){+}g(y_j))^2} .
\label{eq:proof_proportion_form11}
\end{eqnarray}
Given $w_i>w_j$, the inequality \eqref{eq:proof_proportion_form11} contradicts to Eq.\eqref{eq:proof_proportion_form2}.

\noindent \textit{Case 2: $f(x_i){<}g(y_i)$.} Because of $g(y_j){>}g(y_i){>}f(x_i)$, there has  
\begin{eqnarray}
\frac{f'(x_i)g(y_i)}{(f(x_i){+}g(y_i))^2}  {>} \frac{f'(x_i)g(y_j)}{(f(x_i){+}g(y_j))^2} {>} \frac{f'(x_j)g(y_j)}{(f(x_j){+}g(y_j))^2},
\label{eq:proof_proportion_form12}
\end{eqnarray}
which contradicts to Eq.\eqref{eq:proof_proportion_form1}.

Therefore, for any two targets $B_i$ and $B_j$ with $w_i>w_j$, there must exist $x_i>x_j$ and $y_i>y_j$. 
This concludes the proof. \done

\end{document}